\documentclass[10pt,journal,compsoc]{IEEEtran}
\ifCLASSOPTIONcompsoc
  \usepackage[nocompress]{cite}
\else
  \usepackage{cite}
\fi
\ifCLASSINFOpdf
   \usepackage[pdftex]{graphicx}
   \graphicspath{../images/}
\else
   \usepackage[dvips]{graphicx}
   \graphicspath{../images/}
\fi
\usepackage{amsmath}
\usepackage{array}

\ifCLASSOPTIONcompsoc
  \usepackage[caption=false,font=footnotesize,labelfont=sf,textfont=sf]{subfig}
\else
  \usepackage[caption=false,font=footnotesize]{subfig}
\fi
\usepackage{hyperref}

\usepackage{eurosym}
\usepackage{multirow}
\usepackage[table]{xcolor}
\usepackage{amsthm}

\begin{document}
%
\title{QuIS: The Question of Intelligent Site Selection}

%
%

\author{Sebastian Baumbach,
		Florian Sachs, 
		Sheraz Ahmed, and
		Andreas Dengel
	\IEEEcompsocitemizethanks{
		\IEEEcompsocthanksitem Sebastian Baumbach, Sheraz Ahmed, and Andreas Dengel are with German Research Center for Artificial Intelligence, Kaiserslautern, Germany.
		\IEEEcompsocthanksitem Sebastian Baumbach, and Andreas Dengel are with University of Kaiserslautern, Germany.
		\IEEEcompsocthanksitem Florian Sachs is with University of Magdeburg, Faculty of Economics and Management, Magdeburg, Germany. \protect\\
		Email: sebastian.baumbach@dfki.de
		}
}

%
%

\markboth{Journal of \LaTeX\ Class Files,~Vol.~14, No.~8, August~2015}%
{Shell \MakeLowercase{\textit{et al.}}: Bare Demo of IEEEtran.cls for Computer Society Journals}
%



\IEEEtitleabstractindextext{%
\begin{abstract}
Site selection is one of the most important decisions to be made by companies. Such a decision depends on various factors of sites, including socio-economic, geographical, ecological, as well as specific requirements of companies. The existing approaches for site selection are manual, subjective, and not scalable. The paper presents the new approach \textit{QuIS} for site selection, which is automatic, scalable, and more effective than existing state-of-the-art methods. It impartially finds suitables site based on analyzing decisive data of all location factors in both time and space. Another highlight of the proposed method is that the recommendations are supported by explanations, i.e., why something was suggested. To assess the effectiveness of the presented method, a case study on site selection of supermarkets in Germany is performed using real data of more than $200$ location factors for $11.162$ sites. Evaluation results show that there is a big coverage ($86.4 \%$) between the sites of existing supermarkets selected by economists and the sites recommended by the presented method. In addition, the method also recommends many sites ($328$) where it is benefial to open a new supermarket. Furthermore, new decisive location factors are revealed, which have an impact on the existence of supermarkets. 
\end{abstract}

}

\maketitle

\IEEEdisplaynontitleabstractindextext

%
\IEEEpeerreviewmaketitle

\IEEEraisesectionheading{\section{Introduction}\label{sec:introduction}}

\IEEEPARstart{S}{electing} a facility location is a constitutive investment decision of critical importance for every company that wants to operate a successful business \cite{Strotmann2007, Bhatnagar2005443}. Enterprises face such a decision problem at least once in their lifetime. Negative influences of a site's location cannot be compensated by other factors or actions \cite{woratschek2000, glatte2015location}. Numerous geographical, social, political, economic, or socio-economic factors for each site are usually be taken into account by economists and compared to the specific requirements of a company~\cite{blair1987major}. 
\newline
In the area of site selection, research work traces back nearly a century \cite{lehr1885mathematische, mcmillan1965manufacturers, weber1909urber}. However, due to globalization and digitization, geographical search space as well as amount of data has enormously increased. This data covers not only detailed maps like Google Maps or OpenStreetMap, but also data about demographical and social attributes of inhabitants (for instance, offered by the Federal Statistical Office of Germany as open data). In addition, there exist many companies focusing on the creation of complex and commercial datasets, ranging from regional purchasing power to consumer behavior of citizens. This exponential growth of decisive data makes it increasingly difficult for decision-makers and experts to analyze all relevant information manually. This is why locations are selected top -- down (i.e. from state to region) in practice. As a consequence, some regions are excluded subjectively by experts and the selection is commonly based on subjective criteria \cite{mcmillan1965manufacturers, luder1983unternehmerische, Zelenovic2003}. The remaining sites are weighted with respect to their descriptive attributes and the "best" site wins this election. However, this weighting and selection are highly biased and often based on the individual experience and assessment of the involved experts~\cite{olbert1976standortentscheidungsprozess, schober1990strukturierte, greiner1997standortbewertung, rikalovic2015fuzzy}. As a result, the location decision is not consistently predicated on all information.
\newline
The aim of this paper is to objectively find a suitable site based on analyzing decisive data of all location factors. An attempt is made to build a data driven quantitative model for site selection, where all locations as well as the companies' requirements are treated respectively. Thus, sites are described by their attributes, e.g. \textit{purchasing power}, \textit{number of inhabitants} or \textit{proximity to suppliers}, which are called location factors in economics. This approach makes it necessary to collect data of all sites together with all of their location factors and integrate it into a huge dataset. Companies' requirements are expressed by a combination of these location factors and appropriate weights. Following this approach, a suitable site is a location where its corresponding location factors fulfill the requirements of a specific company. 
\newline
This paper presents the novel approach \textit{QuIS} for data-driven site selection, which analyzes location factors and company requirements. The presented method first builds a data model for which it captures all location factors by incorporating geospatial-temporal data from different sources. This model is structured in geographical, temporal, and hierarchical dimensions and combined with an enhanced recommendation system, where user requirements (constraints and preferences) are matched to the corresponding data by the use of means-end relations. The proposed system is intuitive, flexible, easy to use, and only requires knowledge about the company itself. The flexible and intuitive nature of \textit{QuIS} allows it to be used in many different industries with diverse requirements. The existing approaches for site selection are based on complex decision support systems that are mostly refused by the industry as these models tend to be too complex, cumbersome and confusing~\cite{massmann2007kapazitierte}. To resolve this issue, the presented method not only provides the recommendation but also provides detailed explanation about how a decision was made. 
In particular, this paper made following contributions:

\begin{itemize}
	\item A generic, hierarchical data model for geospatial-temporal data. With this model, arbitrary homogeneous geospatial data as well as their changes over time can be integrated and queried according to the three dimensions (space, time and features). It can be used not only for site selection but for all applications, which benefit from geospatial data like crisis prediction.
	\item A large dataset consisting up of more than $200$ location factors for all $11,260$ municipalities of Germany. The existing datasets used in literature are quite small and contain only limited information i.e., either all required location factors or all available municipalities of a nation, but not both. 
	\item An extended recommendation system capable of dealing with large geospatial-temporal data for the purpose of site selection. 
	\item A case study about German's supermarkets where the actual locations of $1,700$ supermarkets in Germany are compared with the recommendations of the site selection system. The results showed that there is a big overlap ($86.4 \%$) between existing stores selected by the expert and sites recommended by the proposed methods. 
	\item An explanation engine facilitating decision makers with explanation on decision, i.e., why a specific site was selected. 
	\item New location factors, which have an impact on the suitability of sites for supermarkets. Unknown patterns behind locations and its characteristics are identified by analyzing sites and its location factors in-depth. These results are used to overcome the cold start problem of the proposed site recommender system.
\end{itemize}

Evaluation results of the case study shows that the each supermarket chain follows its own strategy and prefers different sites in Germany for their stores.

The rest of this paper is organized as follows. In \textit{Section \ref{sec:background}}, an introduction to site selection and background information are provided. \textit{Section \ref{sec:start_of_the_art}} summarizes and assesses the state-of-the-art for site analysis in economics, maths and computer science as well as locational datasets. \textit{Section \ref{sec:methodology}} presents the proposed approach for data-driven site selection, whereas \textit{Section \ref{section:data_model_dataset}} depicts the corresponding hierarchical data model and the dataset used to perform the evaluation. \textit{Section \ref{sec:case_study}} presents the conducted case study about supermarkets in order to demonstrate the viability of \textit{QuIS}. \textit{Section \ref{sec:analyzing_lf}} introduces the in-depth analysis for identifying hidden, influencing location factors and Section \ref{sec:Explanation_engine} describes the explanation engine. Finally, the results are discussed in \textit{Section \ref{sec:conclusion}} together with a summary and an outlook for future work.

\section{Site Selection: An Overview}
\label{sec:background}

The \textit{site} or \textit{location} of a company is the geographic place where a company does business \cite{luder1983unternehmerische}. The location characteristics, which are relevant for the operating performance, are called \textit{location factors} \cite{weber1909urber}. While the economic viewpoint examines the global distribution of companies from a broader perspective, the  intra-company viewpoint deals with the spatial arrangement of resources within buildings, which is also known as layout planning. In this paper, the focus is on the business perspective of site respective location selection rather than on the economic or intra-company point of view. 

Site selection is mainly done by analyzing and comparing the location factors of different places. These factors are the properties that are influential towards a company's goal achievement \cite{liebmann1969grundlagen}. The important location factors are usually selected by a company based on their own demands and each company might possess different requirement. Table \ref{tab:LocationFactors} shows the most common location factors grouped into 10 major categories \cite{badri2007dimensions}. 


\begin{table}[!ht]
	\def\arraystretch{1.25}
	\caption{Categorized Location Factors, a Selection}
	\label{tab:LocationFactors}
	\centering
	\resizebox{1\columnwidth}{!}{
		\begin{tabular}{l|p{7.5cm}}
			\textbf{Category} & \textbf{Location Factor} \\ 
			\hline
			Transportation           & Highway facilities, railroad facilities, waterway transportation, airway facilities, trucking services, shipping cost of raw material, cost of finished goods transportation, shipping cost of raw material, warehousing and storage facilities.  \\ 
			\hline
			Labor                    & Low cost labor, attitude of workers, managerial labor, skilled labor and wage rates, unskilled labor, unions, and educational level of labor. \\ 
			\hline
			Raw Materials            & Proximity to supplies, availability of raw materials, nearness to component parts, availability of storage facilities for raw materials and components, and location of suppliers. \\
			\hline
			Markets                  & Existing consumer market, existing producer market, potential consumer market, anticipation of growth of markets, marketing services, favorable competitive position, population trends, location of competitors, future expansion opportunities, and size of market. \\ 
			\hline
			Industrial Site          & Accessibility of land, cost of industrial land, developed industrial park, space for future expansion, availability of lending institution, closeness to other industries. \\ 
			\hline
			Utilities                & Attitude of utility agents, water supply, wastewater, cost and quality, disposable facilities of industrial waste, availability of fuels, cost of fuels, availability of electric power, and availability of gas. \\ 
			\hline
			Government Attitude      & Building ordinances, zoning codes, compensation laws, insurance laws, safety inspections and stream pollution laws. \\ 
			\hline
			Tax Structure            & Industrial property tax rates, state corporate tax structure, tax free operations, and state sales tax. \\ 
			\hline
			Climate \& Ecology       & Amount snow fall, percent rain fall, living conditions, relative humidity, monthly average temperature, air pollution, and ecology. \\ 
		\end{tabular}
	}
\end{table}

Location factors can be further divided into hard and soft factors. The first group is quantifiable and has direct impact on the economic viability of locations as they directly influence the cost and revenue calculations. This includes, among others, the number of skilled workers and their education, the proximity to suppliers, and tax rates. The second group contains qualifiable factors, which cannot be integrated into the calculations of the company due to their descriptive and non-numerical nature. This comprises, for instance, political conditions, prestige of sites, or leisure programs. However, the later group is becoming more and more important for site analysis \cite{blair1987major}.
\newline
These criteria are highly dependent on the industry and the specific company. For retail stores, the close proximity (purchasing power, public transport, or prestige) is of crucial importance. In comparison, the availability of labor, infrastructure, and tax considerations are of high interest for factories. Sites for the production of hazardous substances require the lack of residential and other industrial facilities in their close surrounding.

As a consequence, methods for site selection do not only have to take all necessary location factors into account, but also model the specific requirements of companies in their new sites. These requirements, however, differ from company to company.

\section{State-of-the-Art in Locational Datasets and Site Selection Methods}
\label{sec:start_of_the_art}

The research community mainly focuses on the collection of new locational datasets (containing both a list of sites and location factors) as well as the development of algorithms to link these datasets with companies' requirements. Within the last century, economists have published various methods for site selection. In the last decade, approaches originally came from the area computer science were often applied in site analysis. So far, recommender systems, however, have not been used for site selection, although they were developed to make suggestions to users. The rest of this section summarizes and assesses the current state of research on locational datasets, site analysis in science and industry, and recommender systems.

\subsection{Locational Datasets}
\label{subsection:Locational_Datasets}

Researchers have studied many locational datasets over time. Their investigations are analyzed regarding three criteria, which are I) the spatial granularity (nations, countries or municipalities/cites as area of focus), II) the number of sites, and III) the number of location factors. The ultimate goal is to analyze with a high spatial granularity (municipalities/cites) all existing sites within the area of focus by considering a large number of locational factors. Table \ref{tab:Locational_Datasets} shows the findings.

\begin{table}[!ht]
	\def\arraystretch{1.25}
	\caption{Taxonomy of Locational Datasets used in Site Selection Studies}
	\label{tab:Locational_Datasets}
	\centering
	\resizebox{1\columnwidth}{!}{
		\begin{tabular}{r|r|r|r}
			\textbf{Studies} & \textbf{Number of Location Factors} & \textbf{Number of Sites} & \textbf{Spatial Granularity} \\ 
			\hline
			Assaf et al. \cite{assaf2015attracting} & $23$ & $123$ & Nations \\ 			
			\hline
			Amini \cite{amini2015multi}  & $11$ & $15$ & Districts  \\ 
			\hline
			Arauzo-Carod \cite{arauzo2005determinants}  & $4$ & $720$ & Municipalities \\ 
			\hline
			Aydin et al. \cite{aydin2013gis}  & $18$ & $5$ Provinces & $250m \times 250m$ Cells \\ 
			\hline
			Carver \cite{carver1991integrating}  & $16$ & $5,878$ & $2 \times 2km$ Cells \\  
			\hline
			Donevska et al. \cite{donevska2012regional} & $12$ & $1$ region & Area for landfills \\
			\hline
			Gorsevski et al. \cite{gorsevski2013group} & $7$ & $27$ & Countries \\
			\hline
			Mohajeri and Amin \cite{mohajeri2010railway} & $45$ & $5$ & Cities \\
			\hline
			{\"O}n{\"u}t et al. \cite{onut2010combined} & $8$ & $6$ & City Districts \\
			\hline
			Rikalovic and Cosic \cite{rikalovic2015fuzzy} & $4$ & $45$ & Municipalities \\ 
			\hline
			Vahidnia \cite{vahidnia2009hospital} & $5$ & $5$ & Cities \\
			\hline
			Vasiljevi{\'c} et al. \cite{vasiljevic2012gis} & $14$ & $1$ region & Area for Landfill \\
		\end{tabular}
	}
\end{table}

In conclusion, investigations of sites have only been made on the basis of small datasets with both a small number of location factors and assessed sites. None of the studies has used datasets, which fulfills likewise all three requirements. Mohajeri and Amin \cite{mohajeri2010railway} analyzed the most location factors ($45$) in order to find the best candidate for a railway station, but only evaluated five sites. Arauzo-Carod \cite{arauzo2005determinants} studied $720$ Catalan municipalities, but took only $4$ location factors (population density, skill level of human capital, and infrastructure) into account. Assaf et al. \cite{assaf2015attracting} evaluated international hotels and the location factors that matter most. They collected a dataset for $123$ sites, which contains $23$ location factors (e.g. performance of the hotel industry, crime rates, or tax rates). However, they only focused nations as spatial entity, instead of considering sites with an higher spatial granularity, such as countries or municipalities.

\subsection{Site Selection in Science and Industry}
\label{subsection:SiteSelectionMethods}

In literature, many approaches have been proposed how a site can be selected. In fact, there is no fixed procedure. However, few guidelines and operating procedures exist, with each having its own drawbacks and limitations. In general, the existing methods for site selection used by economists are very complex and time extensive (as they require a lot of manual configuration and analysis), which limit their applicability in practice \cite{Zelenovic2003, woratschek2004dienstleistungsmanagement, arauzo2010empirical}. Most of these methods do not scale with data or request certain constraints (i.e. homogeneous distribution), which are normally not fulfilled in practice. 

\subsubsection*{Hierarchical Analysis}
\label{subsubsection:Hierarchical_analysis}

In literature and practice, the site selection process is often divided into multiple phases where the regional focus is reduced in each phase. Zelenovic split it up into a macro and micro selection \cite{Zelenovic2003}. The first phase addresses the issue of finding the right state while the second one looks for a specific site within the previously chosen state. Bankhofer divided the whole process into four phases of selecting continent, country, municipality and then final location in that order \cite{Bankhofer2001}. However, this division "top-down" is impartial and inefficient as it requires manual analysis and selection. The exploration and selection of all decisive location factors by hand is not feasible since the amount of data is increasing exponentially.

\subsubsection*{Regression Analysis}
\label{subsubsection:regression}

Models like Discrete Choice Models \cite{hausman1978conditional} or Count Data Model \cite{mcfadden1974conditional} for weighting and selection of location factors have become increasingly complex over the years \cite{arauzo2010empirical}. They include not only numerous, but also constantly changing factors based on the varying contexts of each model. There is no similar study, which reveals identical findings, but suggests different subsets of location factors \cite{blair1987major}. Even though these models are theoretically helpful, they are too complex and therefore, time and cost intensive in a real scenario. This leads managers to base their decisions not only on given facts, but rather on their personal and emotional judgment \cite{olbert1976standortentscheidungsprozess, schober1990strukturierte, greiner1997standortbewertung, rikalovic2015fuzzy}.

\subsubsection*{Methods Used by Economists}
\label{subsubsection:economic}

Woratschek and Pastowski studied different methods in economics used for location selection, namely \textit{Checklist Methods}, \textit{Selection by Elimination} and \textit{Scoring Models} \cite{woratschek2004dienstleistungsmanagement}. These "conventional approaches" utilize managers' judgment in terms of their knowledge and experiences. This is the reason why these methods should only provide a set of systematic steps for problem solving since the capabilities and experiences of the analyst significantly influence the final results \cite{kuo2002decision}.

\textit{Checklist Methods} are a very basic way to assist in the selection process. Relevant location factors for a company get listed and weighted by experts for different sites. These weights are the degree how good a location fulfills a given requirement. Table \ref{tab:Checklist} shows an example of a checklist for site selection.

\begin{table}[!ht]
	\def\arraystretch{1.25}
	\caption{Checklist Method, Example.}
	\label{tab:Checklist}
	\centering
	\resizebox{1\columnwidth}{!}{
	\begin{tabular}{l|c|c|c}
		\textbf{Location Factors}	& \textbf{Location 1} & \textbf{Location 2} & \textbf{Location 3} \\
		\hline
		Availability of Resources & + & + & o \\
		\hline
		Income Structure & + & - & + \\
		\hline
		Consumer Structure & - & o & o \\
		\hline
		Infrastructure & + & + & + \\
		\hline
		Taxes & o & o & + \\
		\end{tabular}
	}
\end{table}

The advantage of the \textit{Checklist Method} is its simplicity and the little time it needs to be done. However, such a checklist does not provide any in-depths analysis. In addition, the presented results are only of qualitative, but not of quantitative nature.

\textit{Selection by Elimination} is an extension to \textit{Checklist Method}, which immediately removes a location not fulfilling a requirement. At the end of this process, after gradually eliminating all non-sufficient alternatives, only sites that satisfy all given requirements are left for further selections. However, this method has some drawbacks. Since all of the criteria can eliminate a site, it is implicitly assumed that criteria cannot compensate each other. This might lead to the exclusion of sites that are just below an expected quality of one requirement, but are excellent in all other aspects.

\textit{Scoring Models} are another extension of the \textit{Checklist Method} approach, where location factors are weighted and sites are rated by a given user-defined scale (e.g. from 1 to 5).  Those ratings are multiplied by the respective weight given to location factors and summed up in the end. This leads to an actual numerical result and a comparable ranking of locations based on pre-defined factors and their importance. Table \ref{tab:ScoringModel} gives an example.

\begin{table}[!ht]
	\def\arraystretch{1.25}
	\caption{Scoring Models, Example.}
	\label{tab:ScoringModel}
	\centering
	\resizebox{1\columnwidth}{!}{
	\begin{tabular}{l|c|c|c|c|c}
		\multicolumn{2}{|l|}{}                         & \multicolumn{2}{c|}{\textbf{Location A}}      & \multicolumn{2}{c|}{\textbf{Location B}} \\
		\hline
		\textbf{Criteria}         & \textbf{Weighting} & \textbf{Points}       & \textbf{Total Points} & \textbf{Points}  & \textbf{Total Points} \\
		\hline
		Consumer                  & 60 \%              &                       &                       &                  &                       \\
		\hline
		Avg. Income               & 15 \%              & 4                     & 0,6                   & 5                & 0,75                  \\
		No. of inhabitants zone 1 & 25 \%              & 2                     & 0,5                   & 3                & 0,75                  \\
		No. of inhabitants zone 2 & 20 \%              & 2                     & 0,4                   & 4                & 0,8                   \\
		\hline
		Competitors               & 30 \%              &                       &                       &                  &                       \\
		\hline
		No. of competitors        & 10 \%              & 1                     & 0,1                   & 1                & 0,1                   \\
		Market share              & 20 \%              & 5                     & 1                     & 1                & 0,2                   \\
		\hline
		Infrastructure            & 10 \%              &                       &                       &                  &                       \\
		\hline
		Customers                 & 5 \%               & 2                     & 0,1                   & 2                & 0,1                   \\
		Transportation            & 5 \%               & 3                     & 0,15                  & 1                & 0,05                  \\
		\hline
		Result                    & $\Sigma$ 100 \%      & \multicolumn{1}{l|}{} & 2,85                  &                  & 2,75                  \\
	\end{tabular}
	}
\end{table}


\subsubsection*{Multiple-Criteria Decision Analysis}
\label{subsubsection:mcdm}

Site selecting can also be modeled as Multiple-Criteria Decision Analysis (MCDA) problem in which a new facility is located among several alternatives by minimizing or maximizing at least one objective function (e.g. costs, profit, revenue, or distance to suppliers). Farahani et al. have published a detailed survey of these multiple criteria facility location problems \cite{farahani2010multiple}. 

Multiple MCDA techniques have been used for site selecting in literature, such as TOPSIS \cite{lai1994topsis}, ELECTRE \cite{roy1991outranking}, and PROMETHEE \cite{mareschal1984promethee}.  Many studies have been conducted in order to rank alternative sites, especially in the field of landfill site selection \cite{donevska2012regional, vasiljevic2012gis}, environmental problems \cite{salminen1998comparing, cheng2002using, marinoni2005stochastic}, or renewable energy systems \cite{pedrero2011application, gorsevski2013group, aydin2013gis, uyan2013gis}. 
\newline
Saaty \cite{thomas1980analytic} introduced the Analytical Hierarchy Process (AHP), which is one of the most important method in selecting optimized alternatives \cite{vahidnia2009hospital}. The limitation of AHP is that it does not allow the items (here, location factors) of the hierarchical model to have dependences on each other. To overcome these drawbacks, the Analytic Network Process (ANP) \cite{saaty2001decision} was proposed modeling "the decision process as a sequence of uni-directional, hierarchical relationships rather than complex networks of objectives" \cite{tuzkaya2008analytic}. ANP allows independence interaction and feedback between items. 

MCDA methods for site selecting focus merely on the analysis of location criteria and their ranking by experts, rather than on the investigation of criteria and their actual impact on the business performance of sites in practice. A detailed analysis of advantages and disadvantages of these MCDA methods has been published by Velasquez and Hester \cite{velasquez2013analysis}. 


\subsection{Recommender Systems as Site Selection}
\label{subsection:RecommenderSystem}

The existing recommender systems commonly used for suggesting movies \cite{lee2010collaborative}, books \cite{carrer2012social}, or music \cite{nunez2012implicit} cannot be used for site selection. This is because these systems cannot be configured to incorporate an application context with a complexity similar to location analysis. Recommender systems mainly operate in the two-dimensional $User \times Item$ space. Traditionally, items are defined by their attributes, which contain keywords or numeric values. Users are described by their preferences basically expressing their interest in certain items. A detailed introduction to recommender systems can be found in \cite{nextGen}. However, this kind of model is not suitable for recommending sites for three reasons.

For suggesting suitable sites to companies, different forms of knowledge have to be employed, which is more than just interest on specific items. These \textit{knowledge dimensions} cover preferences, constraints, context, and domain knowledge about the users' needs, about other users and the locations themselves as well as past recommendations \cite{felfernig2008constraint}. As a consequence, companies' requirements and sites cannot be modeled with the traditional concepts of users respective items as they cannot completely hold the aforementioned knowledge dimensions. 

Furthermore, this focus of recommender systems to make suggestions based upon information about users and items lead to another disadvantage. These systems do not incorporate additional contextual information beyond the two dimensions $User \times Item$, which is crucial in this scenario. However, an extended recommender system with the capability of \textit{multidimensionality} can utilize supplemental dimensions to enhance the suggestions. Evident examples are aggregated users' demands extracted from social media, information about the hierarchical structure of locations ("Munich is a city in Bavaria, which is a state in Germany"), or temporal data as the location factors are naturally changing over time \cite{adomavicius2005incorporating}.

Finally, most recommender systems deal with single-criterion ratings, such as ratings of movies. In site selection, however, it is essential to integrate \textit{multi-criteria ratings}, e.g., a company can have different constraints, which must be fulfilled, as well as multiple preferences, which should be satisfied \cite{nextGen}. These kind of constraints cannot be employed in recommender systems as the utilized profiles just model the users' interest in items. They do not contain any information about the degree of fulfillment, whether a criteria is mandatory, preferable, or optimal to fulfill.

\section{QuIS: Data-driven Site Selection} 
\label{sec:methodology}

\begin{figure}[!ht]
	\centering
	\includegraphics[width=\columnwidth]{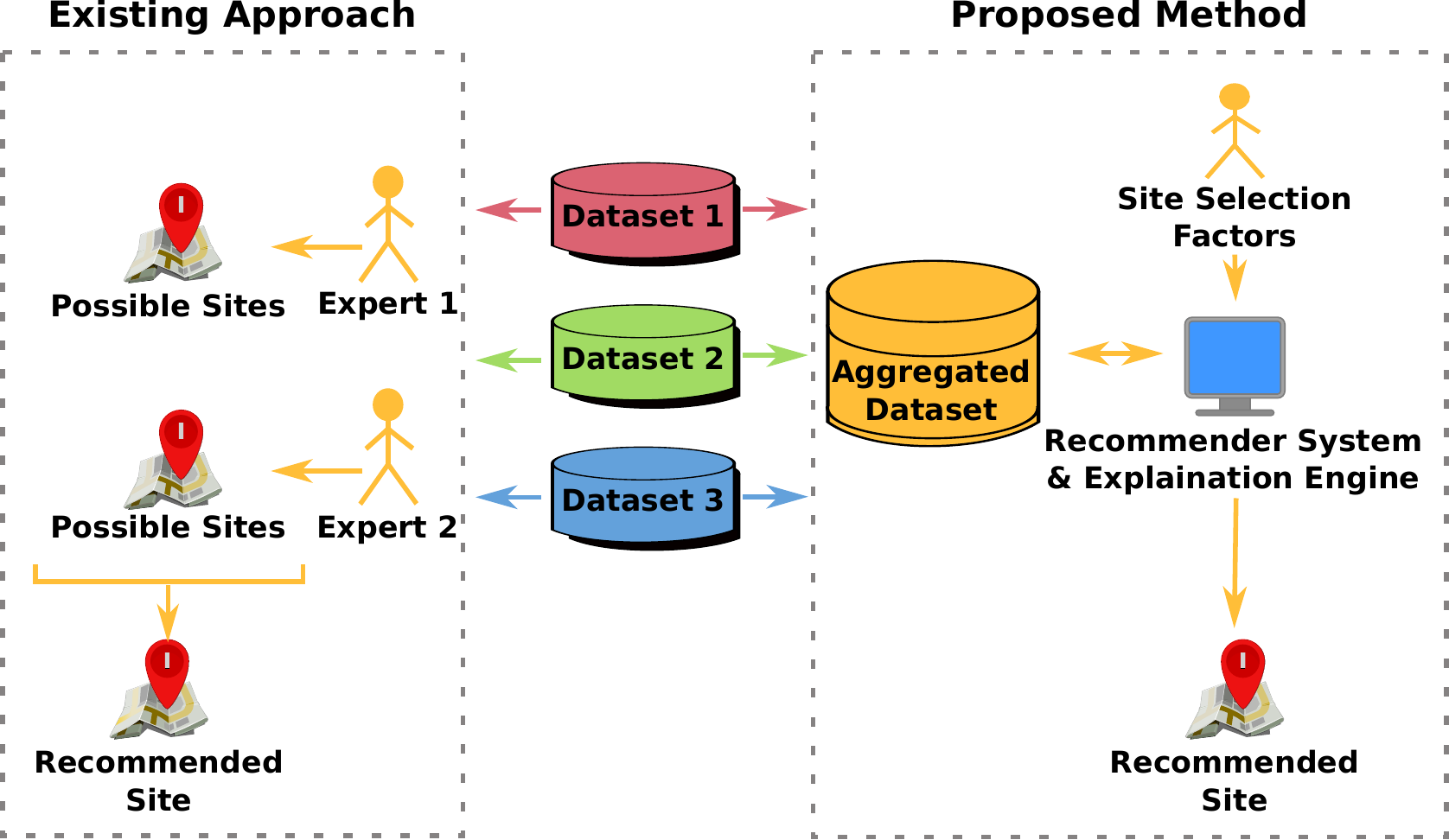}
	\caption{Site Selection: Existing vs. proposed Site Selection.}
	\label{fig:Comparison_Approaches}
\end{figure}

The proposed approach for site selection is based on a novel hierarchical data model and an enhanced version of recommender system. \textit{QuIS} translates companies' requirements for site selection into their location factors and simultaneously, utilizes the available knowledge concerning locations and companies' requirements along with the data model. It turns the \emph{manual exploration of possible sites by decision-makers and experts} into a \emph{definition of requirements by decision-makers \& getting recommendations by the expert system}. Figure~\ref{fig:Comparison_Approaches} illustrates this comparison between traditional and data-driven site selection.

\begin{figure}[htbp]
	\centering
	
	\includegraphics[width=0.9\columnwidth]{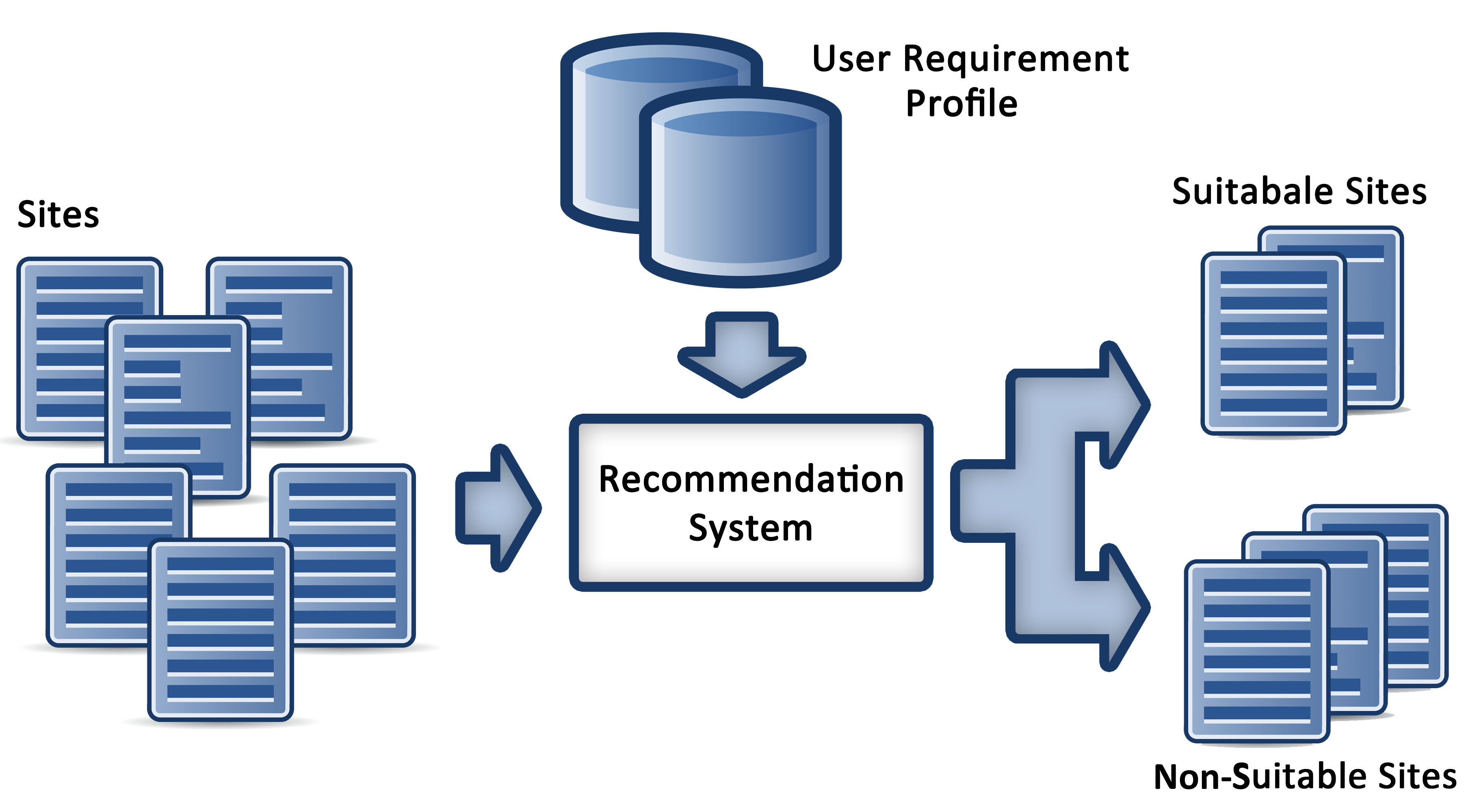}
	\caption{Illustration of the Recommender System Idea}
	\label{fig:SystemIdea}
\end{figure}

Site selection is a task that is executed with low frequency, which means that there is no chance to develop a user profile based on past transactions. Additionally, companies do not usually provide feedback regarding the profitability of their locations in detail. Hence, the common methods used in content based and collaborative approaches cannot be applied here, as they are highly dependent on either user profiles or rated items respectively. 
\newline
A strategy to deal with such cold start problems is user questioners to manually create an initial profile. \textit{QuIS} however, employs a semi-automatic approach to create these profiles (hereafter called \textit{User Requirement Profiles}). They are built by the help of the users respective companies. Section \ref{sec:analyzing_lf} introduces an semi-automatic approach for creating these user profiles. 
\newline
Once the user profiles are defined, the recommendation system utilizes these profiles to identify the suitable sites. The actual recommendation task is defined as a \textit{constraint satisfaction problem}. Figure \ref{fig:SystemIdea} provides an overview of the presented method where the recommender system uses \textit{User Requirement Profiles} to separate suitable sites from non-suitable sites.

\subsection{User Requirement Profile}
\label{subsection:User_Profile}

\begin{figure}[htbp]
	\centering
	\includegraphics[width=\columnwidth]{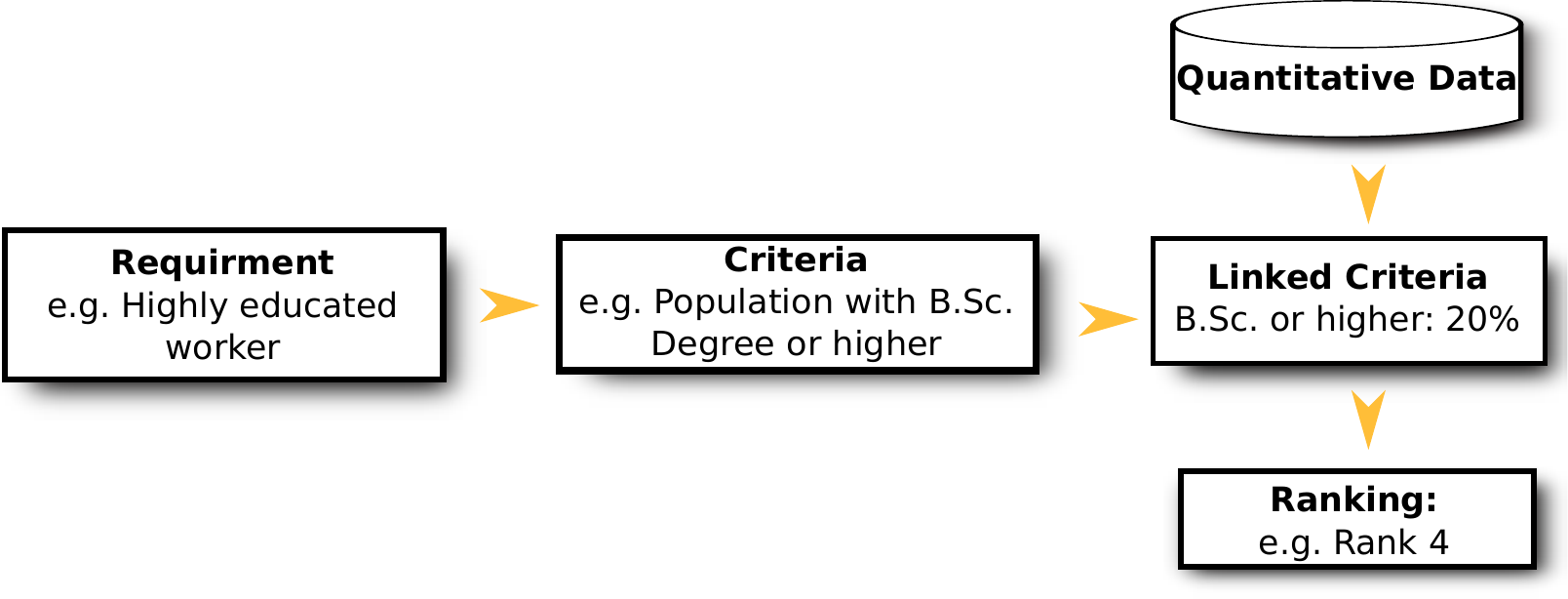}
	\caption{Conceptual View on the proposed User Requirements Profile.}
	\label{fig:RequirementProcess}
\end{figure}

The most important problem in site selection is that decisions should be made based on qualitative measurement and evaluation. However, data available is mostly quantitative. Inspired by the recommendation methods used by economists (see \cite{woratschek2004dienstleistungsmanagement}), all requirements are concretized to criteria and linked to quantitative data, given by location factors, in the first step. Afterwards, these criteria are weighted based on their relevance for the company before they are used for ranking sites. Figure \ref{fig:RequirementProcess} illustrates this concept behind the use of \textit{User Requirements Profiles}.

\begin{figure}[htbp]
	\centering
	\includegraphics[width=0.8\columnwidth]{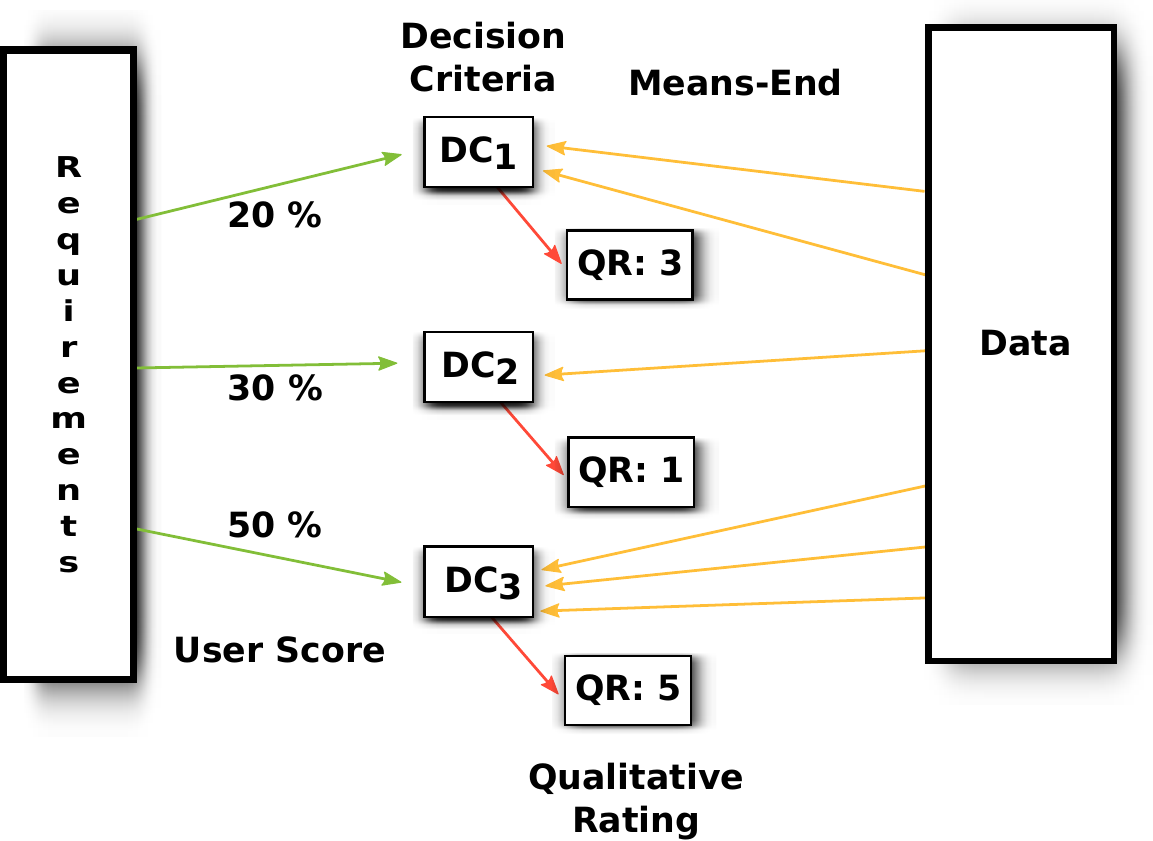}
	\caption{Abstraction of the URP Model.}
	\label{fig:URP_Abstraction}
\end{figure}

The \textit{User Requirements Profile} is a composition of \textit{Decision Criteria}, \textit{Qualitative Ratings}, and \textit{Means-End Assignments}. Figure \ref{fig:URP_Abstraction} shows the workflow for creating \textit{User Requirements Profile}. These profiles incorporate ideas and concepts from Felfernig and Burke in which constraints, preferences and means-ends are applied in a recommender system \cite{felfernig2008constraint}. \textit{Decision Criterion} represents a company's requirements for a new location (green). This requirement description is then linked to the actual location factors using \textit{Means-End Assignment} (orange). \textit{Qualitative Rating} is the evaluation of a criterion based on the quantitative data given by the location factor (red) and thus, used to make qualitative suggestions. 

\subsubsection*{Decision Criteria}
\label{subsubsection:DC}

\textit{Decision Criteria} are the top-level components of \textit{User Requirements Model}, representing abstract requirements, such as \textit{good infrastructure} or \textit{weak competitors}. A set of multiple criteria together makes up a single user profile. The importance of each criterion is reflected by an associated weight. \textit{Decision Criteria} also utilize the concept of 'Scoring and Weighting' of location factors originally introduced by scoring models in economics. (see \cite{woratschek2004dienstleistungsmanagement}). It simply combines the idea of 'Criteria' and 'Weighting' (see Table \ref{tab:ScoringModel}).

Furthermore, the concept of 'Selection by Elimination' is used where all criteria are ranked based on their importance and locations get eliminated if they do not fulfill all of the requirements. In contrast, in the proposed system each criterion can be marked independently from each other \cite{felfernig2008constraint}. \textit{Decision Criteria} can either be classified as 'Must-Have' (representing an elimination condition if this criterion is not fulfilled) or as constraint in form of 'Preferences' or respectively 'Nice-To-Have'. This adjustment helps to accurately represent user requirements.

Using the concepts of 'Scoring and Weighting' and 'Selection by Elimination' has the advantage of being already approved in practice as it takes two established commonly used methods and combining their individual strengths while leaving out their disadvantages. Furthermore, it also enables an easy transition from the existing manual site selection to the proposed approach. Since requirements for site selection are already known by experts, these requirements can easily be rebuilt within the proposed site evaluation model without any need to restructure or remodeling them. 

\subsubsection*{Means-End Relation of Requirements and Location Factors}
\label{subsubsection:QR}

Each \textit{Decision Criterion} are evaluated based on the given location factors. However, companies' needs cannot always be directly translated to location factors. In most cases, requirements represented by \textit{Decision Criteria} cannot be derived from a single factor. To resolve this problem, the presented approach uses fuzzification by the use of \textit{Means-End Assignments}~\cite{cao2007MeansEnd}. A location factor might satisfy different quality needs and in return a qualitative need encompasses multiple attributes. The proposed approach utilizes means-end relations incorporating domain knowledge, which allows reasoning about "how particular items (the means) satisfy particular needs or requirements of the user (the ends)"~\cite{felfernig2008constraint}.

The purpose of \textit{Means-End Assignments} is to associate a \textit{Decision Criterion} with $n$ location factors. As a consequence, companies' requirements are translated to multiple location factors.

\textit{User Requirements Profile} uses weights to describe the importance of each \textit{Decision Criterion} and thus, specifying the association between the profile and all of its criteria. In case of the \textit{Decision Criteria} and \textit{Means-End Assignments}, however, the means-end can be expressed in any mathematical formula to build up \textit{Decision Criterion}. In the following example, four associated means-end, representing four location factors at the end, are combined in the following way:

\begin{equation}
\frac{(ME_{1}~+~ME_{2})^{ME_{3}}}{2*ME_{4}}
\label{eq:ExampleME}
\end{equation}

This way, the utilized \textit{Means-End Assignments} allow flexibility and accuracy when it comes to constructing decision criteria.

\subsection{Qualitative Rating: Evaluating Decision Criteria}
\label{subsection:Evaluating_Requirements}

The \textit{Decision Criterion} is used to compare and rank sites. However, existing algorithms used in recommender systems are commonly used to find items that are similar to ones already rated positively \cite{nextGen}. This is not the case for site recommendation due to the given application context where no user rating is available in advance. Therefore, pre-defined ideal values for \textit{Decision Criteria} are used by the recommendation engine for comparison. Thus, a \textit{Decision Criterion} is the expression of a company's requirement towards a site's attributes in relation to a given ideal value. This way, a degree of fulfillment can be calculated how a good a particular site fulfills the company's requirement. Additionally, such a requirement can be optimally marked as 'mandatory'. 
\newline
The calculation of this ideal value is done by using a bounding box, which can be defined by the user. It expresses a range by specifying both lower and upper bounds for this \textit{Decision Criterion}. The degree of fulfillment is calculated in relation to these bounds. For example, a certain criterion is defined as the number of inhabitants and the corresponding bounds are set to $bound_{low} = 2.000$ and $bound_{up} = 10.000$. Considering a site with a population of $8.500$ $value_{site}$, which lies within these bounds, the degree of fulfillment can be declared as: 
\[
\frac{value_{site} - bound_{low}}{bound_{up} - bound_{low}}
\]
For the above example, the formula results in a score of 0.8. However, this is actually just one possible method for the calculation. To make the system more adaptive to the companies' requirements, following other evaluation strategy were also integrated into the system: 

\begin{itemize}
	\item a boolean strategy that evaluates whether a value is within the bounds set,
	\item strategies that use exponential or logarithmic formulas to calculate scores,
	\item different handling of values that are outside of the bounds set.
\end{itemize}

Thresholds are used to give users respective companies the possibility to specify the degree of fulfillment, which is considered by them to be sufficient. \textit{Means-End Assignments} expressed in arbitrary mathematical formula provides the required flexibility to meet the user's needs as the exact way of comparing and calculating a score heavily depend on the specific requirement. For instance, a retail store is aiming at the maximization of the average purchasing power whereas the crime rates should score better the lower the actual values are. 
\newline
To sum up, the proposed system offers the following features regarding \textit{Decision Criteria}:

\begin{itemize}
	\item Defining compare values in a meaningful way.
	\item Marking them as 'mandatory' with a necessary degree of fulfillment.
	\item Combine multiple location factors in a flexible and mathematical way.
	\item Enable varying means of evaluation.
\end{itemize}

\subsection{Calculating Recommendations}
\label{subsection:Constraint_based}

Ultimately, the recommendation task is defined as a \textit{constraint satisfaction problem}, which has been adapted to site selection. The solution is a list of sites, such that this list is consistent with the requirements of the company as well as the location factors of sites. 

Given a set of the company's requirements, the recommendation can be calculated by using a constraint-based recommender $R_{constr}$, which computes solutions \{$site_i \in AllSites$\} for a given site recommendation task.

\newtheorem{mydef}{Definition}

\begin{mydef}
	\label{def:CSP}
	Based on \cite{felfernig2008constraint}, a site recommendation task is defined as Constraint Satisfaction Problem $(C, S, CR \cup COMP \cup FILT \cup SITES)$ where $C$ is a finite set of variables representing potential requirements of the company and $S$ is a set of variables defining the basic properties of the sites. Furthermore, $CR$ is a set of the company's \textit{Decision Criteria},	$COMP$ represents a set of (incompatibility) constraints, $FILT$ is a set of filter constraints, and $SITES$ specifies the set of offered sites. 
\end{mydef}

A solution to a given site recommendation task $(C, S, CR \cup COMP \cup FILT \cup SITES)$ is a complete assignment to the variables of $(C, S)$ such that this assignment is consistent with the constraints in $(CR \cup COMP \cup FILT \cup SITES)$. A weighting can be defined on those relations to fine tune the relevance of every location factor according the companies' \textit{Decision Criteria}.

\section{Data Model and Dataset}
\label{section:data_model_dataset}

In order to rank sites according the companies' requirements, \textit{QuIS} needs to be connected with a data model capturing sites as well as location factors. The data model incorporates geospatial-temporal data from various sources and structures it in a geographical, temporal, and hierarchical way. 

\subsection{Hierarchical Data Model}
\label{subsec:data_model}

The data model organizes all available sites in a hierarchy according to a nation's federalization. 

\begin{figure}[htbp]
	\centering
	\includegraphics[width=0.9\columnwidth]{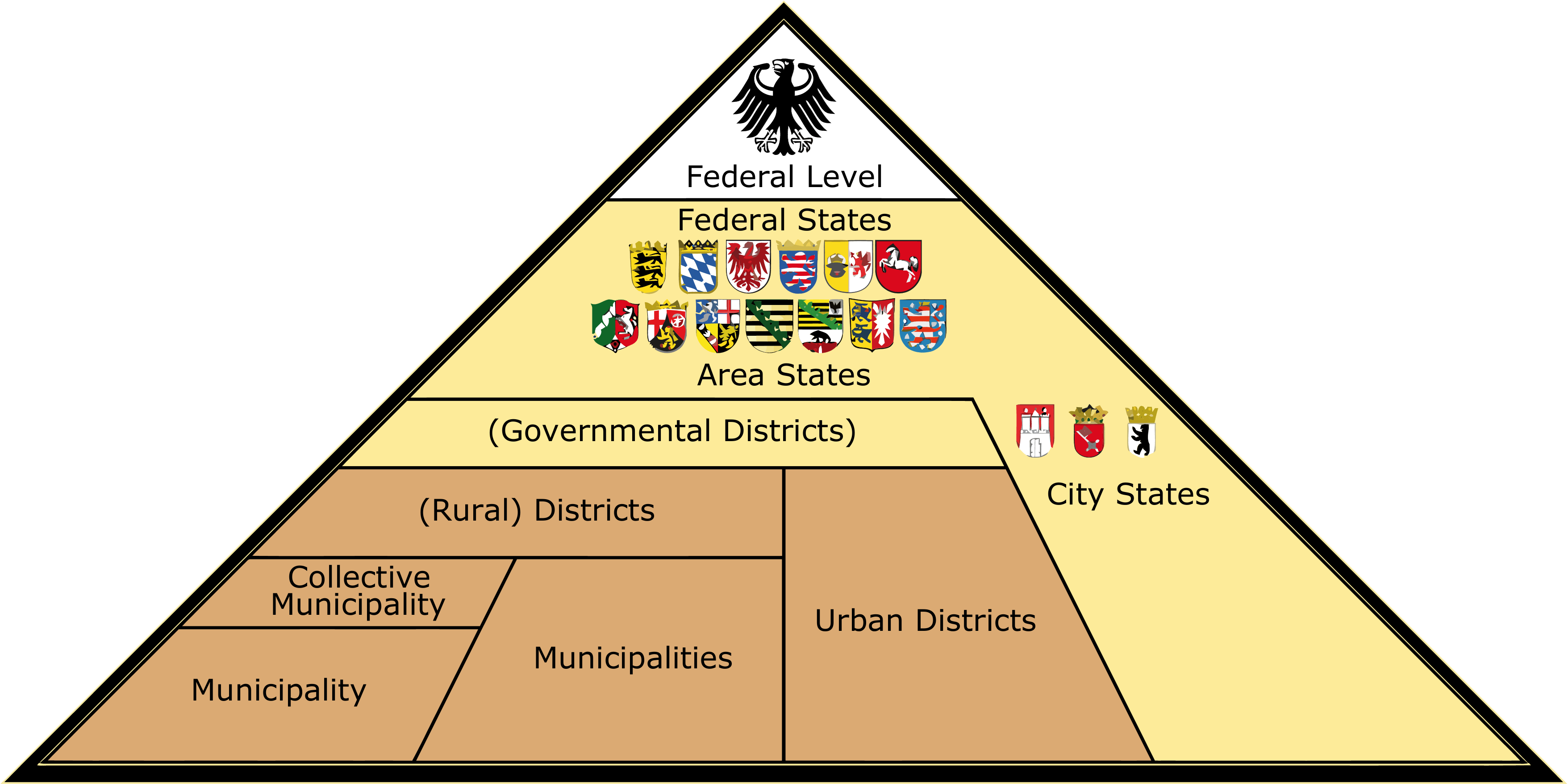}
	\caption[Administrative divisions of Germany.]{Administration Hierarchy Germany, By David Liuzzo \href{http://creativecommons.org/licenses/by-sa/2.0}{CC BY-SA 2.0}.}
	\label{fig:hierarchy}
\end{figure}

Figure \ref{fig:hierarchy} shows the structure of the data model: each state, county, district, and municipality is represented as site. Besides these sites, the data model contains time series, which are the technical representation of location factors. A single time series describes one factor and contains multiple time points as shown in the simplified example:

\theoremstyle{definition}
\newtheorem*{exmpTS}{Exampe Time Series}

\begin{exmpTS}
	name: Population; unit: Count;\\data:[2000, 2001, 2002 ... 2013, 2014, 2015];\\siteID: 120645410340; values:[4000, 4500, 4300, ... 6000, 6250, 6300]
\end{exmpTS}



\begin{figure}[!ht]
	\centering
	\includegraphics[width=\columnwidth, trim=2cm 6cm 2cm 6cm]{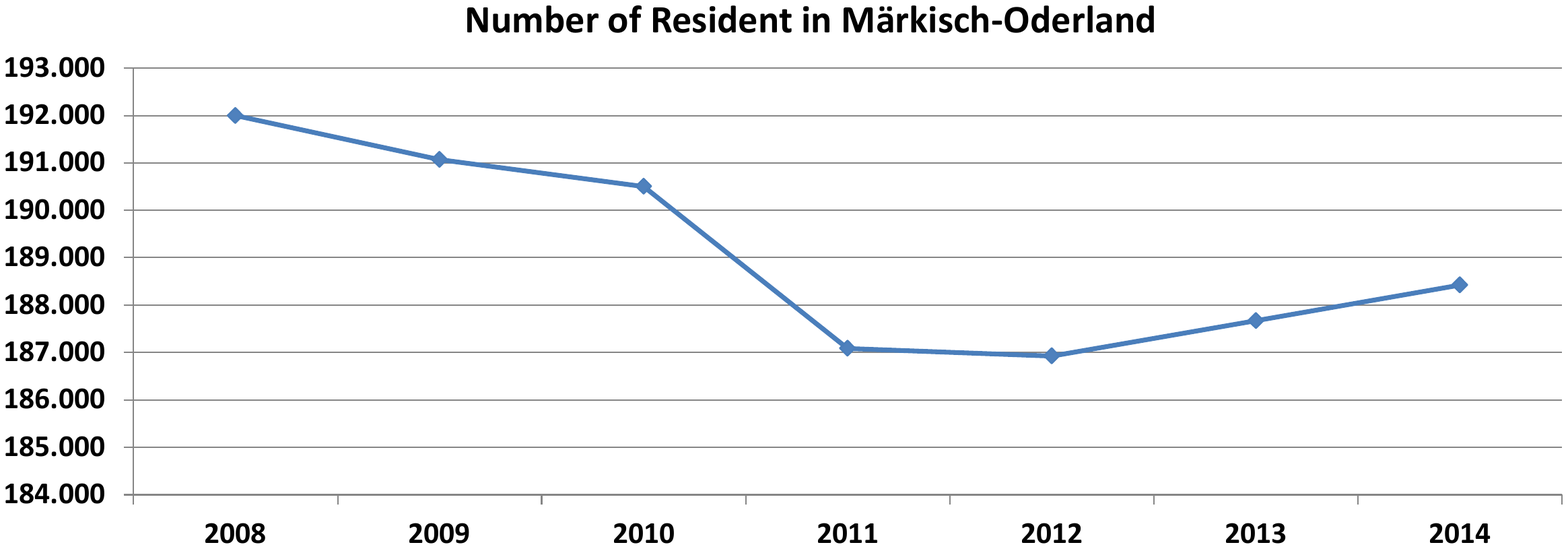}
	\caption{Population for the County \textit{M{\"a}rkisch-Oderland}.}
	\label{fig:ts_Population}
\end{figure}

Figure \ref{fig:ts_Population} shows the time series \textit{population} for the county \textit{M{\"a}rkisch-Oderland}.

Sites in Germany can be identified by using the unique region key officially introduced and employed by the German Government\footnote{See \href{https://en.wikipedia.org/wiki/Community\_Identification\_Number\#Germany}{Wikipedia} for more Information}. The region key is an twelve-digit key that not only uniquely identifies a site in Germany, but also describes the relationships between the four levels (state, county, district, and municipality). The first two digits of the code indicates the state (e.g 12 for \textit{Brandenburg}). The third digit designates the government district, which is deprecated meanwhile, but still exists in the regional key for consistency (e.g. it is 0 in the most cases). The fourth and fifth digit indicates the county (e.g 12 0 64 for \textit{M{\"a}rkisch-Oderland}). The sixth, seventh, eighth, and ninth digit represent the district (e.g. 12 0 64 5410 for \textit{Amt Neuhardenberg}). The last three digits describe the municipality (e.g 12 0 64 5410 340 for \textit{Neuhardenberg}).

\[
Region Key = \underbrace{\overbrace{\underbrace{\overbrace{12}^{State} 0\ 64}_{County} 5410}^{District} 340}_{Municipality}
\] 

For a location factor (for instance, population or purchasing power) to be stored in the data model, the corresponding time series are stored as a geo-referenced time series in the data model.  
	
Due to the hierarchical structure of the data model, it is possible to aggregate values along the hierarchy in both directions: upwards and downwards. The population of a state for example is the sum of the inhabitants of all regions residual in that state (upwards). Information, which describes large geographic regions in general (e.g. languages or dialects), are linked to the highest level (e.g. states and counties) and get inherited top-down along the hierarchy (downwards). The inheritance of location factors allows storing information only on the lowest level of the hierarchy and by this, reduces data redundancy. 

The separation of sites, location factors, and its values give the necessary flexibility to update, chance, or adopt the data model. New location factors or geospatial information can be added easily. Such use case might be, for example, radio network coverage or resource availability. Changes in or extensions of the hierarchy are also possible to realize.

\subsection{Dataset}
\label{subsec:datamodel}

\textit{QuIS} is evaluated on real data using location factors collected from the German Federal Statistical Office. This data is available for all municipalities and cities of Germany. Figure \ref{fig:hierarchy} gives an overview of the administration hierarchy in Germany. This hierarchy contains Germany with its $16$ states, $402$ counties, $4,520$ districts, and $11,162$ municipalities as well as information about the hierarchy. As for now, municipalities are the lowest level of territorial division. This information is provided by the \href{https://www.destatis.de/DE/ZahlenFakten/LaenderRegionen/Regionales/Gemeindeverzeichnis/Administrativ/AdministrativeUebersicht.html;jsessionid=89930EA06D623C492E12165EBAD494FD.cae4}{Federal Statistical Office of Germany} and can be used without any restrictions and free of charge. 

Data for over $200$ location factors have been integrated into the data model, which is also provided by the \href{https://www.regionalstatistik.de}{Federal Statistical Office}. Available location factors cover topics, such as net income, purchasing power, various information about inhabitants \& population, employees \& unemployed persons, education, land costs, households in number \& size, and companies itemized in size, number, profit, industry, \& employed people. Table \ref{tab:berlin} provides an example for Berlin with a selection of most decisive location factors\footnote{The location factors have been chosen based on the results of Section \ref{subsec:attribute_selection}.}. Ultimately, there are $200$ location factors in total containing data for the last $15$ to $20$ years (depending on the specific location factor) and for all sites (the sum of all states, counties, districts, and municipalities). 

\begin{table}[!ht]
	\def\arraystretch{1.25}
	\caption{Selected Location Factors for Berlin.}
	\label{tab:berlin}
	\centering
	\resizebox{1\columnwidth}{!}{
		\begin{tabular}{p{6cm}|r}
			\textbf{Factor} & \textbf{Value} \\ 
			\hline
			Avg. GDP per Employee & \EUR{71,209} \\ 
			\hline
			Available income per Inhabitant & \EUR{22,586} \\ 
			\hline
			Inhabitant & 3,484,995 \\ 
			\hline
			Employment Rate & 78.9 \% \\
		\end{tabular}
	}
\end{table}

\begin{figure}[!ht]
	\centering
	\includegraphics[width=\columnwidth]{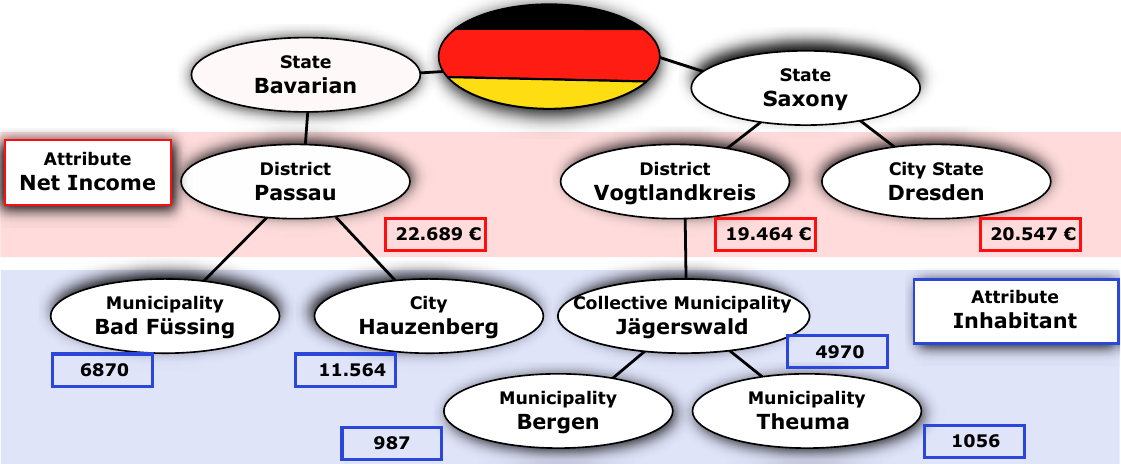}
	\caption{Hierarchical Data Model with Location Factors.}
	\label{fig:overview_data_modul}
\end{figure}

Most of the location factors are available for districts and some have been created on the granularity level of municipalities.  
Each location factor (e.g. net income) exits for all sites within the specific hierarchy level it has been created for (e.g. districts). Location factors are inherent up- and downwards: the inhabitants of a district (blue) are the sum of inhabitants of all municipalities whereas the average net income (red) for a district is also assigned to all of its municipalities and cities. In case of the location factor \textit{net income}, there are $402$ time series in the data model, one for each of the $402$ counties. Figure \ref{fig:overview_data_modul} illustrates these examples.


As stated in Section \ref{subsec:data_model}, each location factor do not only cover values for the current moment, but also for the history. Thus, each location factor technically represents a time series and stores data for the past $15$ to $20$ years. For this study, the latest values for the current year 2016 have been used.

\section{Evaluation: A case study of Supermarkets}
\label{sec:case_study}


To evaluate the effectiveness of the presented approach, a case study on supermarkets is conducted. This case study first focus on site selection for well known supermarkets in Germany and then compare these recommended sites with the sites selected by human experts. The relevant location factors have been provided by experts \cite{greiner1997standortbewertung}. However, \textit{QuIS} is generic and can be applied to datasets collected from any country as well as to a wide range of problems, e.g., gas stations, (fast food) restaurants, cinemas, gold or oil mines.
\newline
Besides evaluating the performance of the proposed recommender system, the collected dataset (see Section \ref{subsec:datamodel}) was analyzed in order to find differentiators for sites with a supermarket and sites where a hypermarket is located. What are the strategies behind companies' decision when they open a new supermarket?
\newline
For effective comparison, the study contains the following supermarket chains \textit{Marktkauf}, \textit{E-Center}, \textit{Edeka Supermarket}, \textit{E-Reichelt}, \textit{E-Aktiv} discounter, \textit{NP Niedrig Preis} and \textit{Lidl}. The regional focus of this study is defined by the sphere of influence of \textit{NP}. This compasses Lower Saxony, Bremen, Saxony-Anhalt, Berlin, Brandenburg, and East-Westphalia, a region of Northrhine-Westphalia. In total, this area contains $1,704$ municipalities respective cities. Table \ref{tab:MunDistribution} shows their distribution across the different states. All $200$ location factors for these sites were taken into account, which includes, among others, purchasing power, population, purchasing power in total, per person and as index, and the average available income per household. Table \ref{tab:CrawledDistribution} shows the number of stores within the area of study (according to their annual report \cite{EdekaUnternehmensbericht2011}).

\begin{table}[!ht]
	\def\arraystretch{1.25}
	\caption{Municipality Distribution}
	\label{tab:MunDistribution}
	\centering
	\resizebox{1\columnwidth}{!}{
		\begin{tabular}{c|c|c|c|c|c}
			Brandenburg & East-Westphalia & Lower Saxony & Saxony-Anhalt & Bremen & Berlin \\
			\hline
			417         & 70           & 996           & 218            & 2      & 1      \\
		\end{tabular}
	}
\end{table}



Also has to be noted that these stores belong to different types of retailing categories. These categories are explained below as well as an attribution of the chosen retailers.

\begin{itemize}
	\item \textbf{Supermarkets} historically are the 'default store'. They sell groceries and other goods of daily needs. Their assortment is of medium size and includes brand-name products as well as home brands. In general, the sales floor of a supermarket is around $1.000m^{2}$ - $2.000m^{2}$. From the retailers listed above \textbf{Edeka} is categorized as supermarket.
	\item \textbf{Discounters}, in comparison to supermarkets, are more dense and focus on lower prices. In general their assortment is simpler and smaller. They focus on fast selling goods with a large percentage of home brands. Because of this, their sales floor is smaller. \textbf{Lidl} and \textbf{NP} are commonly considered as discounter.
	\item \textbf{Hypermarkets} (SB-Warenhaus in German) are characterized by their large sales floor of at least $5.000m^{2}$ and a huge assortment of all different kinds of goods. They also include counter for meat, cheese and fish. What separates them the most are their department for non-food goods, such as electronics, clothing or toys for example. \textbf{E-Center} are categorized as such a hypermarket.
\end{itemize}

\begin{table}[!ht]
	\def\arraystretch{1.25}
	\caption{Crawled Retailer Distribution}
	\label{tab:CrawledDistribution}
	\centering
	\resizebox{1\columnwidth}{!}{
		\begin{tabular}{c|c|c|c|c}
			Supermarket & Lidl & E-Center & Edeka & Niedrig Preis \\
			\hline
			Stores  & 556  & 90       & 589   & 321            \\
			\hline
			Locations & 454	 & 90 			& 480		& 256 \\
		\end{tabular}
	}
\end{table}

\subsection{Site Selection by Experts}
\label{subsec:Profile_Modeling}

\begin{figure}[htbp]
	\centering
	\includegraphics[width=0.75\columnwidth]{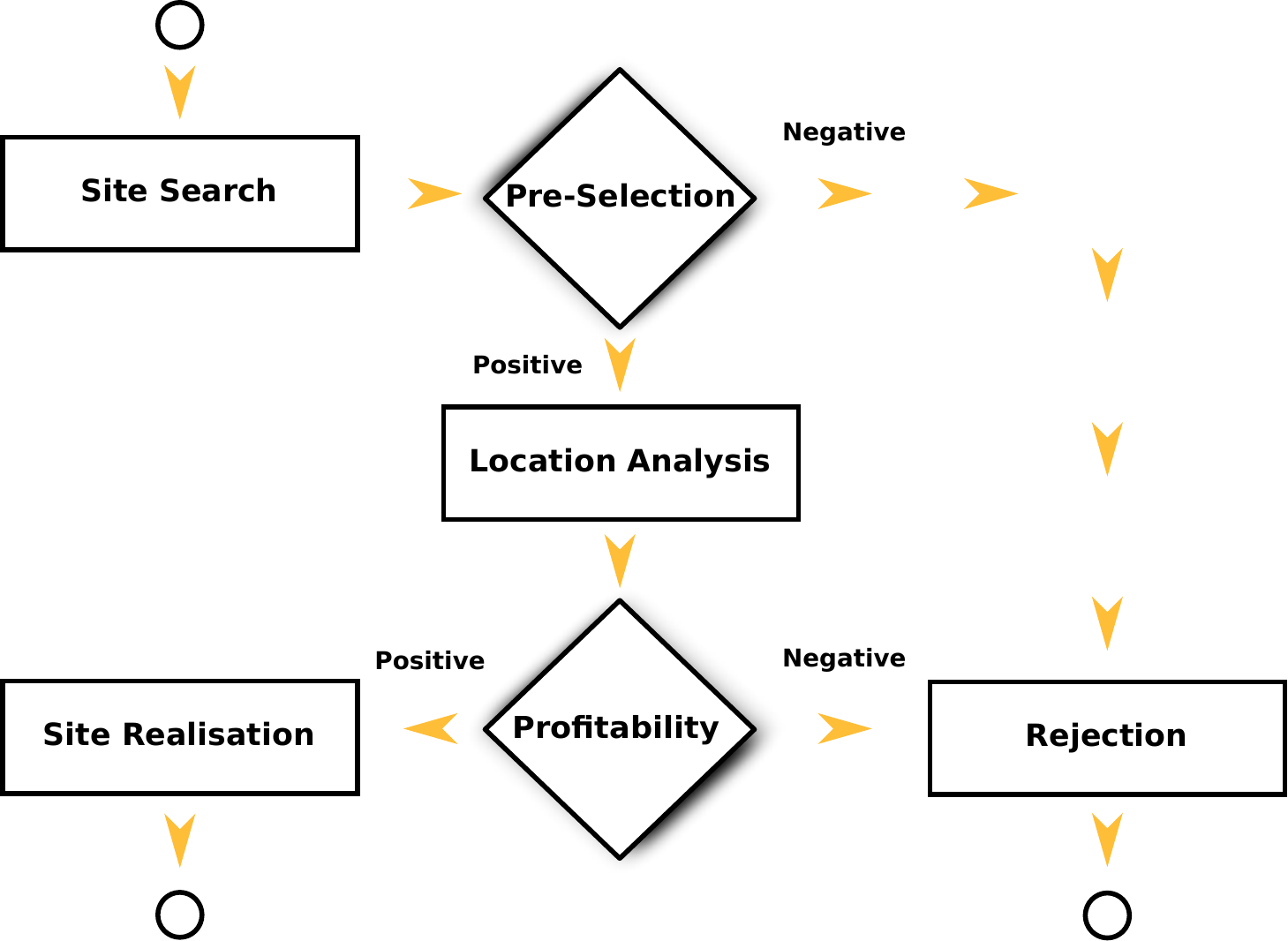}
	\caption{Retailing Site Selection Process}
	\label{fig:SiteSelectionAblauf}
\end{figure}

Greiner, in charge of location selection in the \emph{REWE} Group analyzed in depth how this decision is generally performed and which location factors are of interest for \emph{REWE} supermarkets \cite{greiner1997standortbewertung}. Figure \ref{fig:SiteSelectionAblauf} shows the site selection process, which can be summarized in two phases. 

The macro or pre-selection phase is mostly based on quantifiable numbers. \emph{REWE} mainly uses the location factors \textit{number of inhabitants} in this phase for selection municipalities in Germany. As for Edeka and Lidl, they both list core \& urban population as well as sales floor \& plot area as initial requirements for their retailers (see Table \ref{tab:SMLocationReq}). Beside this factor, the availability of parking places as well as real estate of certain size and their rents are other requirements, which is, however, beyond the scope of this paper (see Section \ref{sec:background}). Any potential site that does not pass those criteria gets sorted out.
\newline
In phase two, the remaining sites are analyzed in-depth. Experts directly visit and assess the suitability of the pre-selected places. This phase includes on-site assessments, analyzing the market structure and competitors, defining the catchment area of a potential store as well as estimating sales. This selection is done by checking quantifiable criteria. The resulting sites are finally checked for their profitability; if the conclusion is positive, the site will most likely be bought.
\newline
According to their expansion strategies, \emph{Edeka}, \emph{Lidl}, and \emph{NP} supermarket chains assert to apply the same criteria\footnote{Source: \href{http://www.lidl-immobilien.de/cps/rde/xchg/SID-6DF6A32A-8E7E665B/lidl_ji/hs.xsl/5186.htm}{www.lidl.de} and \href{https://www.edeka-verbund.de/Unternehmen/de/edeka_minden_hannover/immobilien_minden_hannover/expansion_minden_hannover/edeka_expansion_minden_hannover.jsp?}{www.edeka-verbund.de}.}.

\begin{table}[!ht]
	\def\arraystretch{1.25}
	\caption{Location Requirements for Supermarkets}
	\label{tab:SMLocationReq}
	\centering
	\resizebox{1\columnwidth}{!}{
	\begin{tabular}{l|c|c|c|c}
		& Edeka          & E-Center        & Niedrig Preis (NP) & Lidl        \\ 
		\hline
		Core Population      & 5.000          & 10.000          & 2.500              & 5.000       \\ \hline
		Area Population & 8.000          & 25.000          & 5.000              & 10.000      \\ 
		\hline
		Plot Area ($m^{2}$)       & 6.000 - 12.000 & 15.000 - 25.000 & 3.000 - 6.000      & 4.000 +     \\ 
		\hline
		Sales Area ($m^{2}$)      & 850 - 2.300    & 2.300 - 4.500   & 600 - 850          & 800 - 1.400 \\
		\end{tabular}
	}
\end{table}

For the evaluation in this paper, the number of \textit{core and area inhabitants} are used as requirements. Plot and sales area are site object specific parameter and thus, actually not a part of the pre-selecting phase.

\subsection{Site Selection by the Proposed System}
\label{subsec:comparison}

$86.4 \%$ of all locations can be explained by only taking these location factors into account (see Table \ref{tab:Overlap}). The requirements for suitable locations for supermarkets as described in the economic literature have been used to configure the recommender system and its \textit{User Requirement Profile}. This profile contains the \textit{number of inhabitants} and the \textit{regional focus}. \emph{NP} is looking for sites where more than $2,500$ inhabitants live, whereas \emph{Edeka} and \emph{Lidl} search for locations where at least $5,000$ inhabitants are living. \emph{E-Center}, which are stores times larger than the ones mentioned beforehand, are seeking for places with more than $10,000$ inhabitants. Finally, the regional focus has been set to the states Saxony-Anhalt, Brandenburg, and Lower Saxony as the chain \emph{NP} operates in these states only. Within this regional focus, there exist $1,704$ municipalities respectively cities with currently $481$ \emph{Edeka}, $90$ \emph{E-Center}, $256$ \emph{NP}, and $453$ \emph{Lidl} stores (sum $1,280$). 

\begin{table}[!ht]
	\def\arraystretch{1.25}
	\caption{Site Selection : Expert compared to the proposed System.}
	\label{tab:Overlap}
	\centering
	\resizebox{1\columnwidth}{!}{
		\begin{tabular}{cr|c|c|c|c|c}
			\multicolumn{2}{c|}{} & \textbf{Edeka} & \textbf{E-Center} & \textbf{Lidl} & \textbf{NP} & Overall \\
			\hline
			\multicolumn{2}{c|}{No. of existing Supermarkets} & 481 & 90 & 453 & 256 & 1280 \\ 
			\hline
			\multirow{2}{*}{\textbf{Recommended Sites}} & No. & 364 & 80 & 428 & 224 & 1096 \\
			\cline{2-7}
			& Percentage & 74.8 \% & 88.8 \% & 94.5 \% & 87.5 \% & 86.4 \% \\
		\end{tabular}
	}
\end{table}

The lower accuracy of $74.8 \%$ for Edeka can be explained by the fact that Edeka works as a franchise with independent market owners. These stores probably do not take advantage of the full optimization potential, as it is the case for the centrally controlled company \emph{Lidl}.

Additionally, $328$ suitable sites were found where currently no supermarket of competitors is located and which can be recommended. Table \ref{tab:Recommendations} contains these recommendations.

\begin{table}[!ht]
	\def\arraystretch{1.25}
	\caption{Recommended Sites for new Supermarket Stores.}
	\label{tab:Recommendations}
	\centering
	\resizebox{1\columnwidth}{!}{
		\begin{tabular}{cr|c|c|c|c}
			\multicolumn{2}{c|}{} & \textbf{Edeka} & \textbf{E-Center} & \textbf{Lidl} & \textbf{NP} \\
			\hline
			\multirow{2}{*}{\textbf{Recommended Sites}} & Total & 270 & 303 & 412 & 559\\
			& Without Markets & 88 & 49 & 51 & 140 \\
		\end{tabular}
	}
\end{table}

Table \ref{tab:Filialen} shows the distribution of supermarket stores. $94.5 \%$ of the \emph{Lidl}, $74.8 \%$ of \emph{Edeka}, $87.5 \%$ of \emph{NP} and 88.8 \% of \emph{E-Center} stores are located at municipalities, which fulfill the location factor \textit{number of inhabitants}. Only in $25$ ($5.5 \%$) cases, \emph{Lidl} opened a store at a site with less than $5,000$ inhabitants. $23.3 \%$ and $11.1 \%$ of \emph{Edeka} respectively \emph{E-Center} stores can be found at sites, which has less inhabitants than actually required. \emph{NP} has selected in $32$ cases ($12.5 \%$) a municipality with less than $2,500$ inhabitants.

\begin{table}[!ht]
	\def\arraystretch{1.25}
	\caption{Relation of the Supermarket Chains to the No. of Inhabitant.}
	\label{tab:Filialen}
	\centering
	\resizebox{1\columnwidth}{!}{
		\begin{tabular}{cc|p{1.5cm}|p{1.5cm}|c}
			\multicolumn{2}{c|}{\multirow{2}{*}{Store}} & \multicolumn{2}{c|}{\textbf{Criteria Inhabitants Fulfilled}} & \multirow{2}{*}{Overall} \\
			& & yes & no & \\
			\hline
			\multirow{2}{*}{\textbf{Lidl}} & yes & 428 & 25 & 453 \\ 
			& no & 412 & 839 & 1251 \\
			\hline
			\multicolumn{2}{c|}{Overall} & 840 & 864 & 1,704 \\
			\hline
			\hline
			\multirow{2}{*}{\textbf{Edeka}} & yes & 364 & 117 & 481 \\ 
			& no & 270 & 953 & 1223 \\
			\hline
			\multicolumn{2}{c|}{Overall} & 634 & 1,070 & 1,704 \\
			\hline	
			\hline
			\multirow{2}{*}{\textbf{E-Center}} & yes & 80 & 10 & 90 \\ 
			& no & 303 & 1311 & 1614 \\
			\hline
			\multicolumn{2}{c|}{Overall} & 383 & 1,321 & 1,704 \\	
			\hline	
			\hline
			\multirow{2}{*}{\textbf{NP}} & yes & 224 & 32 & 256 \\ 
			& no & 599 & 849 & 1448 \\
			\hline
			\multicolumn{2}{c|}{Overall} & 783 & 881 & 1,704 \\				
		\end{tabular}
	}
\end{table}

Additionally, the quality of these results was measured by calculating \textit{Recall}, \textit{Precision}, and \textit{F$_{\beta}$-Measure}. Table \ref{tab:EvaluationMeasures} gives the result for these measures. For the $F_{\beta}$-Measure, $\beta = 2$ was used, which favors recall over precision. The reason is that the conducted study focused on the distribution of supermarkets between sites in Germany rather then the existence of a supermarket in a particular site.

\begin{table}[!ht]
	\def\arraystretch{1.25}
	\caption{Relation of the Supermarket Chains to the No. of Inhabitant.}
	\label{tab:EvaluationMeasures}
	\centering
	\resizebox{1\columnwidth}{!}{
		\begin{tabular}{ll|c|c|c|c|c}
			&                         & \multicolumn{1}{c|}{\textbf{Edeka}} & \multicolumn{1}{c|}{\textbf{E-Center}} & \multicolumn{1}{c|}{\textbf{Lidl}} & \multicolumn{1}{c|}{\textbf{NP}} & \multicolumn{1}{c|}{\textbf{Total}} \\ \hline
			\multicolumn{2}{l|}{\textbf{\begin{tabular}[c]{@{}l@{}}Number of \\ existing Supermarkets\end{tabular}}} & 481                                 & 90                                     & 453                                & 256                              & 1280                                \\ \hline
			\multicolumn{1}{l}{\multirow{2}{*}{\textbf{Recommended Sites}}}                & Number                  & 364                                 & 80                                     & 428                                & 224                              & 1,096                                \\
			\multicolumn{1}{l}{}                                                           & Recall                  & 75,7\%                              & 88,9\%                                 & 94.5\%                             & 87,5\%                           & 85.6\%                              \\ \hline
			\multicolumn{1}{l}{\multirow{2}{*}{\textbf{Qualified Sites}}}                  & Number                  & 634                                 & 383                                    & 634                                & 823                              & 2,474                                \\
			\multicolumn{1}{l}{}                                                           & Precision               & 57,4\%                              & 20,9\%                                 & 27.2\%                             & 67,5\%                           & 44.3\%                              \\ \hline
			\multicolumn{2}{l|}{\textbf{F$_{2}$-Measure}}                                                                  & 71,1\%                              & 53.8\%                                 & 60.6\%                             & 87,5\%                           & 72.1\%                              \\ 
		\end{tabular}
	}
\end{table}

The recall of the proposed system is high with $85.6\%$ across all supermarkets. Thus, the requirement of \textit{number of inhabitants} can be declared as being precise. Especially \textit{Lidl} has only $25$ locations ($5.5\%$) that do not fulfill their own requirements of at least $5,000$ inhabitants. Compared to recall the precision of the recommendation is lower with an overall $44.3\%$. This, however, is accounted for by the economical reasonability: Just because a site is technically suitable for a store, it does not necessarily mean it is also wise to open it there. The potential growth of a company over time is limited and new stores contribute less in the beginning. Area effects, supply chains and competitors also have to be considered, which is out of the scope of this paper.

\section{Identification of hidden influencing Factors}
\label{sec:analyzing_lf}

Since the location factors, as described by experts and supermarket companies themselves, are precise to a high degree, the follow-up question is: Are there any more decisive location factors, irrespective of whether or not they are consciously used by the companies? 
\newline
For the evaluation, location factors for supermarkets were extracted from sites where supermarket stores are located. The results show that there is not only an additional dependency on the purchasing power, but also different strategies of the supermarket chains where to place their stores.
\newline
As a consequence, these decisive location factors, derived automatically from the data, can be used then to initialize the \textit{User Requirements Profile} accordingly. Thus, users respective companies do not have to particularize all of their requirements from scratch, but only have to fine tune existing profiles regarding their specific needs. 

\subsection{Correlation Analysis of Location Factors}
\label{subsec:attribute_selection}

By means of a correlation analysis\footnote{The correlation is a number between $0$ (no correlation) and $+1$ (high correlation) that measures the degree of association between the number of stores per supermarket chain and the corresponding location factors}, all $200$ location factors have been analyzed regarding their impact on the availability of supermarkets. Table \ref{tab:correlation} presents the location factors with the highest correlation scores. It can be seen that, beside the dependency on the number of inhabitants, the purchasing power has an impact on the existence of supermarkets. However, this dependency is only high for \emph{Edeka} and \emph{Lidl}, which are, additionally, correlating among each other. Furthermore, it can be observed that \emph{NP} neither attach importance to the inhabitants nor the purchasing power in the way \emph{Edeka} and \emph{Lidl} set value on it. 

\begin{table}[!ht]
	\def\arraystretch{1.25}
	\caption{Correlation Matrix for Location Factors. $0$ (No Correlation) and $+1$ (High Correlation)}
	\label{tab:correlation}
	\centering
	\resizebox{1\columnwidth}{!}{
		\begin{tabular}{l|l|l|l|l}
			Attribute & Num. of Edeka & Num. of E-Center & Num. of Lidl & Num. of NP \\
			\hline
			Num. of Edeka Stores & \cellcolor{blue!50} 1.0 & \cellcolor{blue!30} 0.61 & \cellcolor{blue!47} 0.94 & \cellcolor{blue!24} 0.48 \\
			\hline 
			Purchasing Power & \cellcolor{blue!47} 0.95 & \cellcolor{blue!30} 0.61 & \cellcolor{blue!49} 0.99 & \cellcolor{blue!24} 0.47 \\
			\hline
			Num. of Inhabitants & \cellcolor{blue!47} 0.94 & \cellcolor{blue!30} 0.60 & \cellcolor{blue!49} 0.99 & \cellcolor{blue!23} 0.46 \\
			\hline
			Num. of Households & \cellcolor{blue!47} 0.94 & \cellcolor{blue!30} 0.60 & \cellcolor{blue!49} 0.99 & \cellcolor{blue!23} 0.46 \\
			\hline
			Num. of Lidl stores & \cellcolor{blue!47} 0.94 & \cellcolor{blue!30} 0.59 & \cellcolor{blue!50} 1.0 & \cellcolor{blue!22} 0.45 \\
			\hline
			Num. of E-Center & \cellcolor{blue!30} 0.61 & \cellcolor{blue!50} 1.0 & \cellcolor{blue!30} 0.59 & \cellcolor{blue!23} 0.47 \\
			\hline
			Population Density & \cellcolor{blue!25} 0.51 & \cellcolor{blue!16} 0.32 & \cellcolor{blue!30} 0.61 & \cellcolor{blue!16} 0.32 \\
			\hline
			Num. of NP Stores & \cellcolor{blue!24} 0.48 & \cellcolor{blue!23} 0.47 & \cellcolor{blue!22} 0.45 & \cellcolor{blue!50} 1.0 \\
			\hline
			GDP per Inhabitant & \cellcolor{blue!12} 0.23 & \cellcolor{blue!05} 0.11 & \cellcolor{blue!12} 0.24 & \cellcolor{blue!1} 0.09 \\	
		\end{tabular}
	}
\end{table}

\begin{figure}[htbp]
	\centering
	\includegraphics[width=\columnwidth]{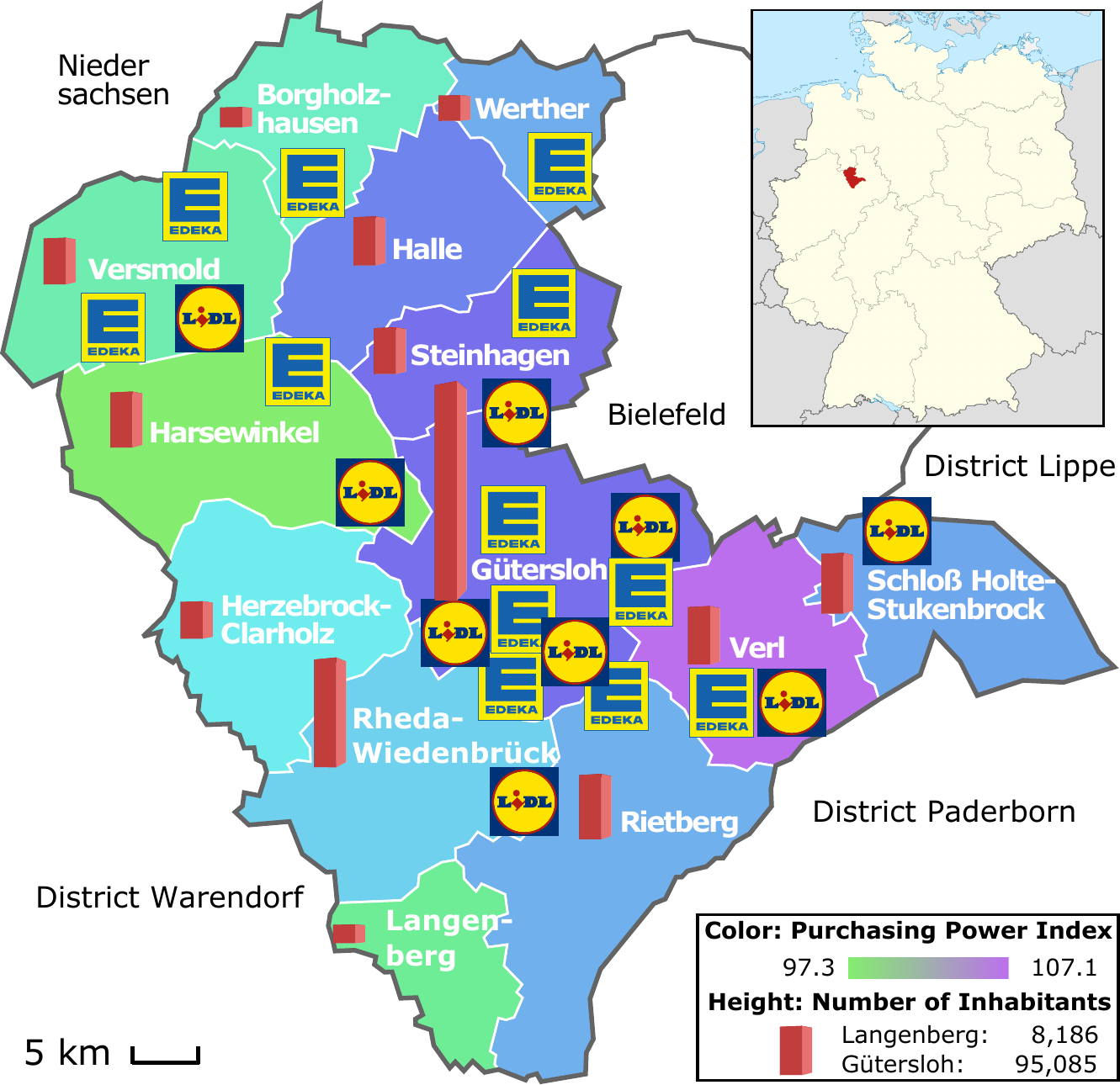}
	\caption{Distribution of Supermarket Locations within the District \emph{G\"utersloh}. Municipalities \& Cities are colored by Purchasing Power Index. Height of Bars visualizes the Numb. of Inhabitants.}
	\label{fig:Municipalities_in_GT}
\end{figure}

Figure \ref{fig:Municipalities_in_GT} gives an example case in detail where the county \emph{G\"utersloh} is visualized together with the number of inhabitants per municipality respective city and their purchasing power index. The district \emph{Herzebrock-Clarholz} has $15,969$ inhabitants and a purchasing power index of 100. This location is among the recommended sites for a new supermarket as it fulfills the requirements. Additionally, there are no supermarkets in the close surrounding available.

\subsection{Revealing the Expansion Strategies of Companies}
\label{subsec:distribution}

Finally, the question is: Why have the supermarket chains chosen these sites among the places, which also suit their requirements? In order to investigate this question, the decisive location factors as calculated by the correlation analysis were used to examine the companies' strategies in detail. For this purpose all locations were grouped into the following ranges based on their population count\footnote{Note that the sample size for cities with more than $100.000$ inhabitants was too small and thus, they were excluded.}:

\begin{enumerate}
	\item 5.000 and below
	\item from 5.001 to 10.000
	\item from 10.001 to 25.000
	\item from 25.001 to 50.000
	\item from 50.001 to 100.000
\end{enumerate}

\begin{figure}[htbp]
	\centering
	\includegraphics[width=\columnwidth]{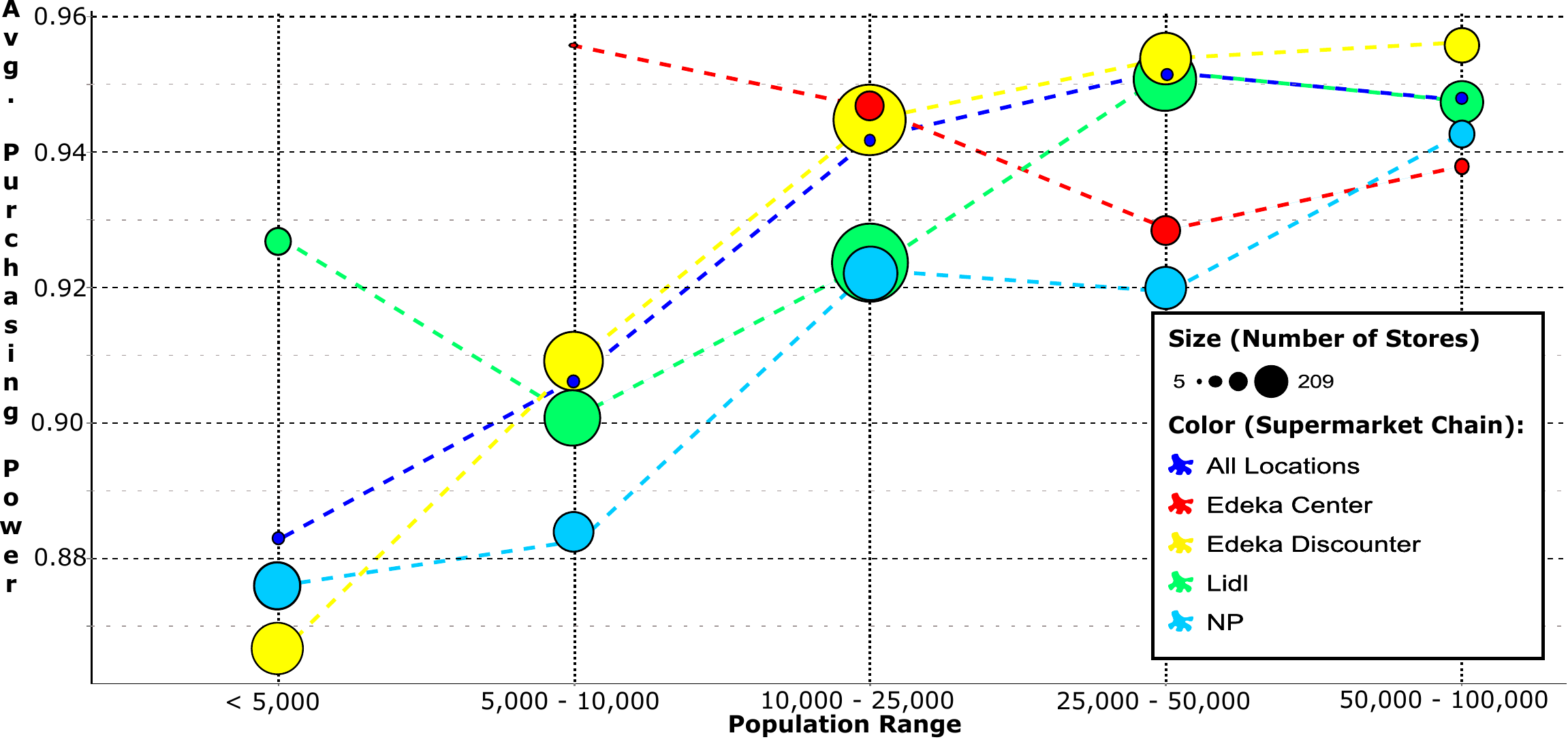}
	\caption{Distribution of Supermarket Chains according to the Average Purchasing Power Index and Grouped by Population.}
	\label{fig:StadtLand}
\end{figure}

Figure \ref{fig:StadtLand} presents the average purchasing power\footnote{The purchasing power index describes the purchasing power of a certain region per inhabitant in comparison to the national average, which itself gets the standard value of 100. If the purchasing power index is $0.84$ for a municipality, these people only possess $84 \%$ of the national average purchasing power.}. This distribution indicates that \emph{NP} mainly focuses on smaller municipalities with lower purchasing power, whereas \emph{Lidl} is concentrated in bigger cities. The strategy of \emph{Edeka} is most likely to be available in all municipalities respective cites with a medium purchasing power as they are mostly avoided by their competitors. The huge \emph{E-Center} stores are mainly located in cities above $10,000$ inhabitants.

\begin{figure}[htbp]
	\centering
	\includegraphics[width=\columnwidth]{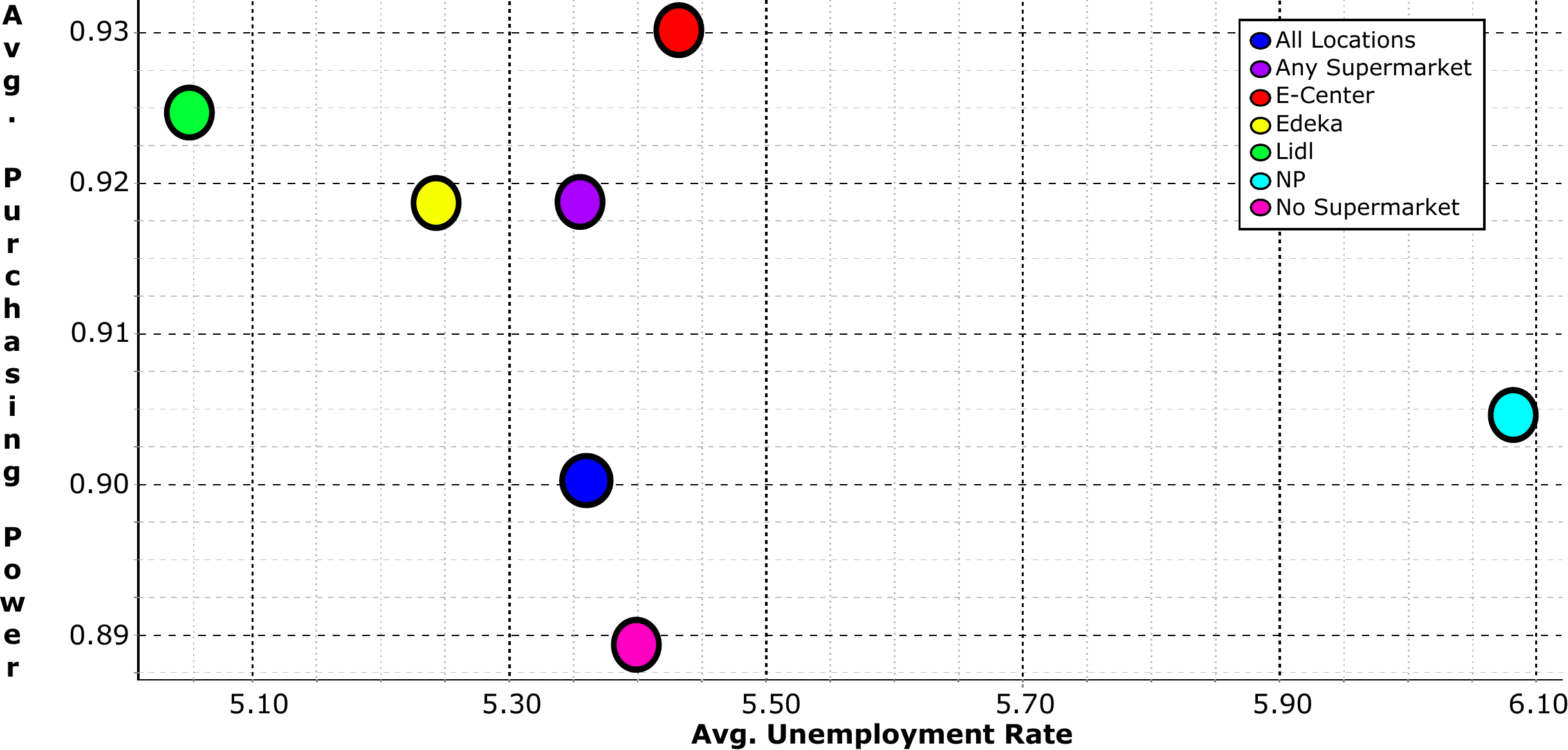}
	\caption{Distribution of Supermarket Chains according to the Average Purchasing Power Index and Unemployment Rate.}
	\label{fig:UnemploymentXPurchasePower}
\end{figure}

Figure \ref{fig:UnemploymentXPurchasePower} shows the average purchasing power index and the average unemployment rate for all municipalities where at least one supermarket of the given chain is located. Thus, the average is presented for \emph{NP} (light blue), \emph{Lidl} (green), \emph{Edeka} (yellow), and \emph{E-Center} (red). As regards the comparison of all possible sites, the distribution for all locations with (purple) and without (pink) supermarkets (regardless of the chain) as well as the all locations (blue) are also presented.
\newline
It can be observed that \emph{Lidl}, \emph{Edeka}, and \emph{E-Center} are focusing locations with higher than average purchasing power and lower than average unemployment rate. In contrast, \emph{NP} opens stores at sites with lower than average purchasing power and higher than average unemployment rate. The Wilcoxon rank sum test (with continuity correction) confirmed that these differences regarding the Purchasing Power Index between Edeka and NP (p = 0.0046) as well as between NP and Lidl (p = 0.0055) are statistically significant. Figure \ref{fig:UnemploymentXPurchasePower} shows the population mean for the purchasing power index, which rank differ for the different supermarket chains and the population ranges. Sites without any supermarket have a low purchasing power, which is most likely why no supermarket has been opened there.
\newline
This can be explained with the two different categories of stores in food retailing: supermarkets and discount stores. Discount stores (e.g. \emph{NP} and to certain extend \emph{Lidl}) aim to sell products at prices lower than normal supermarkets (\emph{Edeka} and \emph{E-Center}) with a focus on price rather than on service, display, or wide choice. Thus, supermarkets have to choose sites with higher purchasing power as they are selling more expensive products than discount stores, which can occupy the market segment of places with lower available money as their products are less expensive.

\section{Explanation engine}
\label{sec:Explanation_engine}

Decision-makers in companies request to understand recommendations made by support systems. If this requirement is not fulfilled as the system acts as a black box, companies will not trust its results and thus, they will also not make crucial business decisions based on these recommendations \cite{massmann2007kapazitierte}. To resolve this issue, \textit{QuIS} not only provides recommendations, but also detailed explanation about how decisions are made. By explaining how the system works and how recommendations are generated, the proposed approach becomes more transparent and trustful.

\begin{figure}[htbp]
	\centering
	\includegraphics[width=\columnwidth]{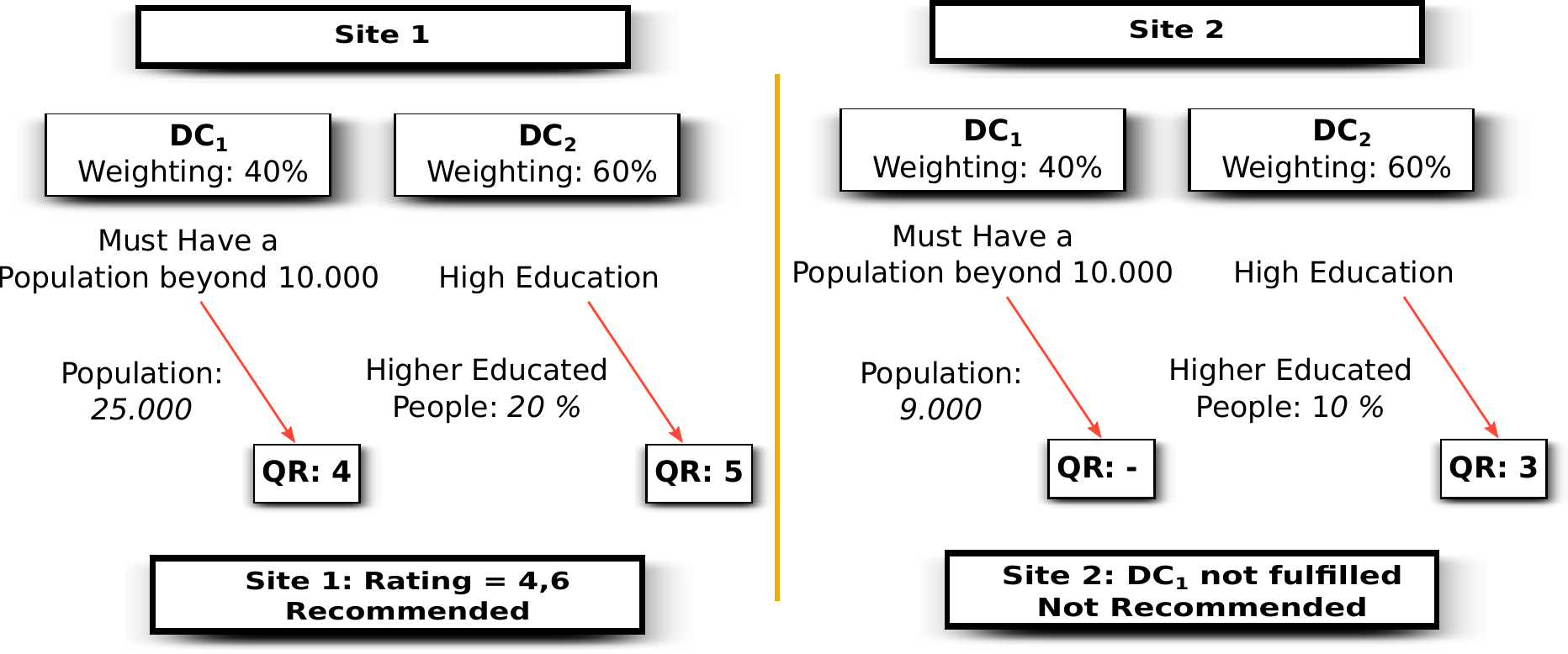}
	\caption{Explaining the Recommendations for two Sites.}
	\label{fig:Explanation_engine}
\end{figure}

The proposed site selection system offers important benefits in terms of the ease with, which the recommendations can be explained and justified. The data model and the \textit{User Requirements Profile} together enable the explanation engine to describe recommended sites according the user's requirements. A site represented by its location factors are compared to the user's requirements and these comparisons are used to describe and explain the recommendations. Figure \ref{fig:Explanation_engine} shows an example.
\newline
In the proposed approach, the data model contains all sites where each site is modeled quantitatively with location factors. The requirements, in turn, are represented with \textit{Decision Criteria}, which are directly linked to location factors by the use of \textit{Means-End Assignments}. Since the weights of all \textit{Decision Criteria} are also given by the respective \textit{Qualitative Ratings}, recommended sites can be described by using the underlying location factors. As the explanation engine knows the user's requirements stored in the \textit{User Requirements Profile}, recommendations can be described with only these location factors, which are relevant to the user. 

Following this procedure, the explanation engine can explain recommendations to users why a site is recommended. For decision-makers, however, it is also of interest why a site is not recommended given alternatives. Sites can be described by the explanation engine with respect to the user's requirement, regardless if this particular site was recommended or not. The aforementioned procedure can be used to explain why certain sites are not recommended.



\section{Discussion and Future Work}
\label{sec:conclusion}

In this paper, the novel approach \textit{QuIS} for data-driven site selection is proposed, which integrates Big Data into the decision making process of companies. It combines a data model with an enhanced recommendation system, which utilizes the existing knowledge in this application context. In contrast to economic methods, the system does not need manual analysis or expert knowledge. Additionally, it is capable of handling all available information about sites. More than $200$ different attributes for all $11,162$ municipalities in Germany have been aggregated and analyzed. The evaluation of supermarkets in Germany shows that there is a big coverage ($86.4 \%$) between existing stores and sites recommended by the proposed methods. Furthermore, \textit{QuIS} also recommends sites ($328$) for supermarket where a store should be opened. Finally, decisive location factors, such as \textit{purchasing power}, were revealed, which have an impact on the existence of supermarkets. 

Most important for the data-driven site selection approach is the dataset. Currently, the level of granularity with municipalities being the smallest unit limits the usability of system. As seen in Section \ref{sec:case_study}, cities with millions of inhabitants (such as Munich and Berlin) are only represented as one site by a singly entity. A more detailed segmentation is needed to generate better recommendations. The reason is neither a conceptual nor technical limitation, but rather a practical issue. In theory, the data model and based upon it, the proposed approach for recommending sites scales in both breadth and depth: other nations or continents can be added as well as a higher granular data (e.g. city districts or streets). In practice, however, it is challenging to find data sources for geospatial data with both a high spatial and temporal granularity.

There is much contextual information that could be incorporated into the recommendation process in the future. Adomavicius and Sankaranarayanan \cite{adomavicius2005incorporating} used a multidimensional approach to incorporate such contextual information. Additional dimensions, can add domain knowledge of locations or contextual information about companies' requirement. Nature reserves are protected areas where construction is not permitted and nuclear power plants had to be kept at a minimum distance to residential areas. Supplementary information might be vital for some companies in their decision process and those additional dimensions help excluding unsuitable locations from the very beginning. 

In the future, most important is the conducting of larger experiments with more location factors, investigated companies, and other industrial sectors such as food chains or shopping malls in order to perform more robust evaluations. Especially companies with hundreds or thousands of stores (e.g. supermarkets, fast food restaurants, or cinemas) are applications of interest for the proposed data-driven site selection.

The application focus can be easily widen as site selection is not limited to industrial companies alone, but governments and private person as well can make use of this system. Especially for private persons however the site hierarchy has to be refined to include smaller area types, such as city districts or even building blocks. Authorities could use the overall recommendation system in reverse and analyze, which kind of companies might be interested in a certain region that leads to another new area for economic research.

The question that is most intriguing is what can be learned from geospatial data. Information about the quality of existing sites will be investigated in order to learn good and poor locations as well as further studies together with economist. The inclusion of qualitative feedback about locations regarding whether it is a 'good' or a 'bad' site will help to further increase the performance. With such information sophisticated machine learning and data mining techniques could be applied to possibly reveal more decisive location factors even unknown to experts in this area. However, such information is not readily available so this approach is most likely only possible in cooperation with one or more companies.

Beside the focus of site selection in this study, the challenge for the future is on pattern recognition and semantic analysis, regardless of the application. In particular, the collected data are not yet further investigated (right now, the temporal aspect is not incorporated) or combined with inferable information like proximity measures. The position of companies, stores and factories, extractable from (linked) geospatial data sources, can also be used to realize data-driven economic competitor and customer analysis in the future.

\ifCLASSOPTIONcompsoc
  \section*{Acknowledgments}
\else
  \section*{Acknowledgment}
\fi

This work was sponsored by Federal Ministry of Education and Research (grant 01IS12050). 

\ifCLASSOPTIONcaptionsoff
  \newpage
\fi

\bibliographystyle{IEEEtran}
\bibliography{./paper}

\end{document}